\documentstyle[11pt,aaspp4,psfig]{article}  

\begin{document}

\def\spose#1{\hbox to 0pt{#1\hss}}
\def\lta{\mathrel{\spose{\lower 3pt\hbox{$\mathchar"218$}}
     \raise 2.0pt\hbox{$\mathchar"13C$}}}
\def\gta{\mathrel{\spose{\lower 3pt\hbox{$\mathchar"218$}}
     \raise 2.0pt\hbox{$\mathchar"13E$}}}
\def\Msun{{\rm M}_\odot}
\def\msun{{\rm M}_\odot}
\def\Rsun{{\rm R}_\odot}
\def\Lsun{{\rm L}_\odot}
\def\half{{1\over2}}
\def\RL{R_{\rm L}}
\def\zs{\zeta_{s}}
\def\zR{\zeta_{\rm R}}
\def\dJJ{{\dot J\over J}}
\def\dMM{{\dot M_2\over M_2}}
\def\tKH{t_{\rm KH}}
\def\eck#1{\left\lbrack #1 \right\rbrack}
\def\rund#1{\left( #1 \right)}
\def\wave#1{\left\lbrace #1 \right\rbrace}
\def\dd{{\rm d}}
\def\u{{\rm u}}

\setlength{\marginparwidth}{1.4truecm}

\title{EVOLUTIONARY EFFECTS OF IRRADIATION IN CATACLYSMIC VARIABLES}
 
\author{P. McCormick \altaffilmark{1} and J. Frank \altaffilmark{2}}
\affil{Department of Physics and Astronomy, Louisiana State
University,   Baton Rouge, LA 70803-4001, USA}

\altaffiltext{1}{cormick@rouge.phys.lsu.edu}
\altaffiltext{2}{and Space Telescope Science Institute,
3700 San Martin Drive,
Baltimore, MD 21218 (frank@rouge.phys.lsu.edu)}

\begin{abstract}
The orbital evolution of cataclysmic variables in which the companion is 
illuminated by a fraction of the accretion luminosity consists of 
irradiation--driven limit cycles on thermal timescales, superimposed on a 
secular evolution toward shorter periods due to
systemic angular momentum losses.
We show that positive orbital period derivatives during bright phases are a 
natural consequence of the expansion of the companion during high mass transfer
phases in the limit cycle. The irradiation 
instability may be enhanced by consequential angular momentum losses $\dot 
J_{\rm CAML}$
accompanying the limit cycle. We investigate the secular evolution of 
cataclysmic binaries under the combined effects of irradiation and 
$\dot J_{\rm CAML}$ and show that faster than secular orbital period
excursions of either sign may occur. 
We discuss whether the mass transfer 
fluctuations that occur during these cycles can account for the observed 
dispersion in disk luminosities or estimated accretion rates at a given 
orbital period. If indeed irradiation--driven 
and CAML--assisted mass transfer fluctuations on 
timescales faster than secular occur, as discussed in this paper, then we
may be able to predict the relative abundances of the
different types of cataclysmic variable at a given orbital period. For example
this mechanism may explain the relative paucity of dwarf novae with respect to 
nova--like variables between 3 and 4 hours.
\end{abstract}

\keywords{accretion, accretion disks ---
	  binaries: close --- cataclysmic variables -- instabilities
}

\section{INTRODUCTION}
\label{Intr}
Cataclysmic variables (CVs) are semi-detached binary stars with orbital periods
of a few hours ($ \lta .5$ days) containing an
accreting white dwarf and a companion or mass donor star which in most cases is 
a main sequence, hydrogen burning star (see Warner 1995, for a comprehensive review 
of CVs; and also Frank, King \& Raine 1992, for an introduction to accretion physics). 
In these binaries the donor
star transfers mass to its compact companion via Roche lobe overflow. 
For this mass transfer to be dynamically and thermally stable 
the companion must be less massive than the white dwarf and systemic angular 
momentum losses are required.
Binary evolution, in the standard CV evolution scenario
(see e.g. King 1988; Warner 1995), is determined by the 
interplay between mass transfer which tends to expand the orbit and systemic 
angular momentum losses which tend to shrink the orbit. Examples of systemic 
angular momentum losses include gravitational quadrupole radiation, as predicted
from general relativity, and magnetic braking, due to mass loss from the 
companion along the star's magnetic field lines, which produces a braking 
torque. The balance between systemic angular momentum losses and the tendency of 
the binary to expand thus usually results in a stable secular mass transfer rate 
whose magnitude is determined by the rate at which angular momentum is lost from 
the binary. The secular effect of these losses is to shrink the orbit and drive 
the binary to ever shorter periods until, late in the evolution, the companion 
becomes partially degenerate and begins to expand. Thus the shortest orbital
period for CVs with hydrogen--rich companions is $P_{\rm min}\sim 75$ min
(Paczy\'nski \& Sienkiewicz 1981; Rappaport, Joss \& Webbink 1982) in agreement 
with the observed period distribution (see e.g. Kolb 1996).
Most CVs are expected to be in
the phase of secular orbital contraction with the possible exception of
some old systems -- the TOADs -- which may have evolved beyond the period 
minimum 
and are faint except when they undergo large amplitude outbursts 
(Howell \& Szkody 1990; Howell, Szkody \& Cannizzo 1995; 
Howell, Rappaport \& Politano 1997).

An obvious and important feature of the observed orbital period distribution 
of CVs, known as the ``period gap", is caused by the relative paucity of 
systems with periods between 2 and 3 hours. 
Currently the best theoretical explanation for this feature is the
interrupted magnetic braking picture 
(Spruit \& Ritter 1983; Rappaport, Verbunt \& Joss 1983).
According to this picture magnetic braking either 
ceases or is drastically reduced near $P_{\rm orb}\approx 3$ hrs 
as the companion becomes fully convective
and the angular momentum loss rate drops suddenly by a factor $\sim 10$.
The natural tendency of the binary to expand under mass transfer then briefly 
wins, the system detaches, and mass transfer halts.
Since the companion has been driven out of thermal equilibrium by the 
mass loss sustained above the gap, the star returns to its thermal equilibrium 
radius and must wait until gravitational radiation grinds the orbit down to
$P_{\rm orb}\approx 2$ hrs when contact is re--established and the binary 
becomes visible again as a CV.

The energy released by accretion in a disk or accretion column is radiated
away and a fraction of it illuminates the side of the companion facing the
white dwarf. The irradiation of the companion can influence the 
above mentioned secular
balance between angular momentum losses and mass transfer and under certain
conditions cause limit cycles. The stability of mass transfer in binaries
with irradiated companions and the nature of the resulting limit cycles
has been the subject of a number of recent papers (King 1995; 
King et al. 1995; 1996; 1997; Ritter, Zhang \& Kolb 1997; 
Hameury \& Ritter 1997). Earlier studies of the 
reaction of the companion to irradiation assumed that the illumination was
either spherically symmetrical and steady (Podsiadlowski 1991; Harpaz \& 
Rappaport 1991, Frank, King \& Lasota 1992) or adopted other approximations 
and treated
closely related situations (Gontikakis \& Hameury 1993; Hameury et al. 1993;
Harpaz \& Rappaport 1995).

The dispersion in estimated mass transfer rates $\dot M_2$ at a given 
orbital period (Patterson 1984) or similarly the dispersion in estimates of 
accretion disk absolute magnitudes at a given binary orbital period 
(Warner 1987; Warner 1995, see Fig. 9.8; Sproats, Howell \& Mason 1996) 
is unlikely to be the result of 
comparable
dispersion in systemic angular momentum loss rates. 
If cataclysmic variables
were driven at such diverse rates at the same orbital period over secular
timescales, the predicted period minimum and the position and width of the 
period gap would be inconsistent with the observed period distribution
(Verbunt 1984, Hameury, King \& Lasota 1989). 
Instead it has been 
proposed that the dispersion in $\dot M_2$ is the result of {\em cyclical
evolution} on timescales too long to be observed directly but too short 
to affect significantly the secular evolution (Warner 1987; King 1995). 
The irradiation--driven 
limit cycles discussed above may well provide a physical mechanism for such
cyclical evolution as proposed by King (1995) and King et al (1996). 
In subsequent papers, it was shown that 
pure irradiation cycles are limited to orbital periods $\gta 6$ hrs
because the thermal inertia of the convective envelope of the
companion increases rapidly toward lower masses (King et al 1996; 
Ritter et al. 1997). King et al. (1996) noted that a positive
feedback between thermal limit cycles and angular momentum losses
(e.g. by coupled fluctuations in the rate of magnetic braking) would
extend the range of the instability to lower masses and shorter orbital 
periods.

This paper explores the above scenario in detail using numerical simulations.
The effects of cyclical changes in the mass transfer rate on the observable 
orbital parameters of compact binaries are investigated under various 
assumptions including both irradiation  and coupled fluctuations 
in angular momentum losses. Increasing the systemic losses in 
proportion to the mass transfer during a thermal cycle is effectively a 
form of consequential angular momentum loss (CAML, Webbink 1985) and affects 
the overall stability of the binary (King \& Kolb 1995).

\section{BIPOLYTROPIC CODE}
\label{Bipo}
In a bipolytropic binary evolution code the structure of the donor star is 
represented by a composite polytropic model. Much of the following 
draws heavily on work done by Chandrasekhar (1939), Rappaport, Joss, \& 
Webbink (1982), Rappaport, Verbunt, \& Joss (1983) and Kolb  \& Ritter (1992).
In the polytropic model pressure $P$ is taken to be only a function of density 
$\rho$:
	      \begin{equation}
		P=K \rho^{\gamma}=K \rho^{(n+1)/n}  \; ,
	    \label{Bipo.1}
	    \end{equation}  
where $K$ is constant. 
The equation of state  is represented as
	     \begin{equation}
	      P={k \rho T D \over m_{\rm u} \mu} \; ,
	      \label{Bipo.2}
	     \end{equation}
where $k$ is the Boltzmann's constant, $m_\u$ is the atomic mass unit, $\mu$
is the mean molecular weight, and $D$ is an electron degeneracy factor (ratio 
of total gas pressure to ideal gas pressure).
The bipolytropic model divides the lower main sequence star (stars with
masses less than 1.2 solar mass) into an inner radiative region 
(``1'' or ``core") with a polytropic
index of 1.5...4, surrounded by a convective region (``2'' or ``envelope") 
with
polytropic index 1.5.  The boundary of these two regions is determined
by a parameter $Q$, which is the ratio of the radiative region's radius
to the total stellar radius, so that $Q=0$ is 
a fully convective star
and $Q=1$ is a fully radiative star.  The free parameters $n_1$ (the polytropic
index in the radiative region) and $h_1$  (the entropy jump between the top of
the radiative region and the surface of the star) are adjusted to closely
match theoretical and observed lower main sequence radii and luminosities.
   We write the densities and pressure as
\begin{equation}
	 \rho_1 = \rho_{\rm c} {\theta_1}^{n_1}, P_1=
	 K_1 \rho_c^{(n_1+1)/n_1} 
	 {\theta_{1}}^{-(n+1)}
\label{Bipo.3}
\end{equation}
in the radiative region (1), and
\begin{equation}
	 \rho_2 = \rho_{\rm i} {\theta_2}^{3/2}, P_2=K_2 
	 \rho_{\rm i}^{5/3} \theta_2^{5/2} 
	 \label{Bipo.4}
\end{equation} 
in the convective region (2).
Here ``c'' denotes quantities evaluated at the center of the star, ``i''
denotes quantities evaluated at the core/envelope interface, and  $\theta$
is the dimensionless temperature obtained from the Lane-Emden equation  
\begin{equation}
     {1 \over \xi_j^2}{d \over d \xi_j }( \xi_j^2 {d \theta_j \over d \xi_j}) 
     = -\theta_j^{n_j}, j = 1, 2 \; .
\end{equation}   
We solve the Lane-Emden equation
for $\theta$ as a function of $\xi$ (dimensionless radius),with 
$n_1=1.5, 2.0 ...4.0$ in region 1 and $n_2 =1.5$ in region 2. Solutions are 
evaluated at the core/envelope interface, approaching from region 2 and from
region 1, as well as at the stellar surface. These solutions are labeled  
${\rm 2i}$, ${\rm 1i}$, and ${\rm 2s}$ ,respectively. A table of
$\theta_{\rm 1i}$, ${\xi_{\rm 2i}}^2 \theta_{\rm 2i}^\prime$, ${\xi_{\rm 2s}}^2 
\theta_{\rm 2s}^\prime$ and ${(-{\xi_{\rm 2s}}^5 \theta_{\rm 2s})}^{-1/3} $,
where primes denote radial derivatives,
is computed for $\approx 50$ $Q$-values. The above quantities allow one to 
calculate the
pressure and density at the interface, the mass of the radiative region, 
the total mass of the star, and the polytropic constant in the convective region, 
respectively. Our bipolytrope code includes some modest improvements in the 
treatment of the degeneracy, which are important for low masses but is 
otherwise very similar to the version described in more detail in 
Kolb \& Ritter (1992).

In the evolutionary calculations presented here,
the stellar radius, binary separation and
Roche lobe radius all change on timescales comparable to the thermal
timescale of the convective envelope (King et al 1996; 1997). 
Therefore, the stationary mass transfer 
equation does not apply, and we must calculate the mass transfer using
(Ritter 1988; Kolb \& Ritter 1990):
 \begin{equation}
    \dot M_2 = \dot M_{0}\exp 
    {\left({R_2-R_{\rm L} \over H_{P}} \right)} \; ,
  \label{Bipo.5}
\end{equation}
where $H_{P}=\epsilon R_2$ is the pressure scale height, and 
we take $\dot M_0=-10^{-8} \Msun/$yr. The exact value of this constant 
does not matter as we are not interested in the precise depth of contact
which yields the stationary mass transfer. 
In order to reduce trivial numerical instabilities (which can be overcome by smaller
timesteps) we adopt $\epsilon=10^{-3}$
rather than $\epsilon=10^{-4}$, which is more appropriate for near 
main--sequence companions. However, we have done some 
calculations with $\epsilon=10^{-4}$ and find qualitatively similar 
results. Including the smaller $\epsilon=10^{-4}$ results in a higher frequency of 
mass transfer oscillations and a slightly higher peak mass transfer rate. 
In every timestep we calculate the
mass transfer, remove that mass from the secondary, allow it to adjust
obtaining a new $R_2$,
evolve the binary according to the angular momentum and mass losses
experienced, calculate the new $R_{\rm L}$, and iterate.

Using our bipolytrope code we are able to reproduce all of the standard 
evolution of CVs. This includes the existence of a minimum period and a 
period gap between
approximately 2 and 3 hrs, mass transfer rates $\sim 1-2$ orders of 
magnitude larger
above the gap than below, and a ``flag'' when contact is reached at the lower 
limit of the period gap.  A typical example of a  set of standard evolutions  
coming into contact at different periods and converging to a common evolutionary
path is shown in Fig. 1 (see also Stehle, Ritter \& Kolb 1996).

\section{CYCLIC EVOLUTION WITH IRRADIATION} 
\label{Irra}
After a brief discussion of previous related work, we present results of our 
simulations for cyclic evolution driven by irradiation with and without 
taking into account mass loss from the primary during nova explosions.

 Podsiadlowski (1991) showed that in the spherically symmetric irradiation case
with a flux of $F_{\rm irr} \gta 10^{11} - 10^{12}\,{\rm erg}\,{\rm cm}^{-2}\, 
{\rm s}^{-1}$,
a lower main sequence ($M_2 \lta .8\, \msun$) star's radius will 
swell significantly. The expansion 
occurs on the thermal timescale and eventually the star will become fully 
radiative. The irradiated surface acts as a thermal blanket causing the star to lose energy in a less efficient manner. A low mass star under these conditions may expand to a few
times its original equilibrium radius as a
result of the star trying to store the blocked energy in 
its 
gravitational and internal energy. In the completely spherical case this is the 
only 
place the excess energy can be diverted.  In the non-spherically symmetric 
case, 
which is more likely to reflect the true 
nature of the irradiation in binaries, the excess 
energy can be transported efficiently by convection and released through the 
unirradiated side. This has the effect of diminishing the expansion of the star.  
However, even when a fraction of the star's surface
is illuminated, the resultant radial expansion of a few percent is still 
much larger than the atmospheric scale height and may cause mass transfer 
cycles of large amplitude. In our quantitative treatment of this effect we
adopt the simple model introduced by Ritter, Zhang, and Kolb 1997 (hereafter 
RZK)
in which a one zone approximation is used for the superadiabatic layers. In the 
one zone model,
convection is assumed to be adiabatic up to the base of the superadiabatic zone 
($\nabla = \nabla_{\rm ad}$). In the superadiabatic zone, convection is an
ineffective method of energy transport, so the 
simplifying assumption is made that energy is transferred via radiation only
($\nabla = \nabla_{\rm rad}$). 
Irradiation reduces the temperature gradient $\nabla$ by raising the 
photospheric temperature and thus the rate at which energy can be lost
through the superadiabatic zone. In this way the superadiabatic zone acts
as valve which is open when there is no irradiation and restricts the
amount of energy which can be released as the amount of irradiation 
is increased. 
 
In order to describe the effective irradiation 
region we assume the radiation is coming from a point source. 
The irradiation flux at the stellar surface is
\begin{equation}
 F_{\rm irr}(\theta)= {\eta \over 4 \pi} {G M_2 \dot M_2 \over R_1 a^2} h( 
\theta) \; .
\label{Irra.1}
\end{equation}
In the equation above, $\eta$ is a dimensionless 
parameter describing the efficiency of the X--ray
irradiation, $a$ is the binary separation, $R_1$ is the radius of the primary,
and $h(\theta)$ is a geometric factor defined as
\begin{equation}
h(\theta)={\cos{\theta} -f_2 \over (1-2f_2\cos{\theta}+f_2^2)^{3/2}} \; ,
\label{Irra.2}
\end{equation}
where $f_2$ is the ratio of the secondary's radius to the orbital 
separation ($R_2/a$).  We can define a dimensionless irradiating flux as
the ratio of impinging flux over unirradiated stellar flux,
\begin{equation}
   x(\theta)= {F_{\rm irr}(\theta) \over \sigma {T_0}^4}=
   {F_{\rm irr}(\theta) \over F_0} \; ,
\label{Irra.2a}
\end{equation}
and a dimensionless stellar flux as 
\begin{equation}
  G[x(\theta)]= {F \over F_0}=
{{T_{\rm irr}}^4 \over  {T_0}^4}-x(\theta)
\label{Irra.2b}
\end{equation}
the ratio of the emergent stellar flux from an irradiated star to 
that of the unirradiated star.

In this model the effective luminosity of the secondary, taking into
account the effects of external irradiation, is:
\begin{equation}
L=4 \pi {R_2}^2 \sigma {T_0}^4 \left[ {1 \over 2} (1 + f_2(q)) 
+{1 \over 2}\int_0^{\theta_{\rm max}} G[x(\theta)] \sin{\theta} d\theta \right] \; ,
\label{Irra.3}
\end{equation}
where the sum of the bracketed terms gives the fractional effective 
surface area through which the stellar luminosity escapes. The function 
$G(x)$ contains all the information on how the superadiabatic layers 
adjust and how effectively they block the internal luminosity. In this 
paper we present evolutions obtained mostly with the one zone model 
described above, but we also discuss and compare with results of calculations 
done with
more detailed treatments of the blocking (Hameury \& Ritter 1997) in
which full stellar models are used to produce a table of $G(x)$. 

In our first set of irradiated models, shown in Fig. 2, we include 
only the effects of systemic 
angular momentum losses, i.e. gravitational radiation and magnetic braking,
ignoring possible angular momentum losses during nova explosions. 
The mass transfer rate is plotted as a function of 
companion mass rather than orbital period, with progressively larger
$\eta$. This choice makes it easier to compare similar sequences of models
calculated with different assumptions. An inspection of these results shows
that mass transfer rates oscillate only if $\eta \gta .03$ is assumed and 
that they
always die out before the secondary reaches $M_2 =.65\, \msun$. For smaller
values of $\eta$ no mass transfer cycles occur.  These results are in agreement 
with previous analytic and numerical work (King et al 1995; 1996; RZK), but
it is the first time that the evolution has been followed self--consistently
through the period gap and below. These graphs show that significant amounts of
mass transfer oscillations occur when $\eta \gta .05$, and for even larger
values ($\eta \gta .2$) we also get mass transfer cycles 
below the period gap.

In contrast to the models shown Fig. 2, RZK included systemic 
and angular momentum 
losses due to mass loss from the primary assuming all mass accreted was lost in 
nova explosions carrying away the specific angular momentum of the primary. 
We assume for our calculations that mass is transferred in a 
non-conservative 
manner. That is the mass of the compact object remains constant, 
while angular momentum is carried away at a rate 
\begin{equation}         
	   {\dot J_{\rm n.e.} \over J}=  
	   \beta_1 M_2^2/(M M_1)\dMM
\label{Irra.5}
\end{equation}
where $\beta_1$ is an adjustable dimensionless parameter measuring the 
specific angular momentum carried away in units of the specific angular
momentum of the white dwarf. Clearly this angular momentum loss 
is a form of CAML.
In a ``real" CV things are a bit more
complex: between nova outbursts mass is conserved, but the mass accumulated
on the primary is lost on timescales generally much shorter than
those characteristic of the cycle. Thus only in an average sense 
do orbital
parameters evolve following equation $(\ref{Irra.5}) $. 
A more sophisticated calculation akin to the ``hibernation scenario"
(Livio \& Shara 1987; Kovetz, Prialnik \& Shara 1988) where nova 
explosions and the the effects of the irradiation by the hot white dwarf are modeled 
as well is beyond our present scope.
   
A sequence of evolutions with different $\eta$ values and $\beta_1=1.0$, 
that is assuming all the mass loss from the system leaves with the 
specific angular momentum of the primary, are presented in Fig. 3.
These cases are analogous to
the cases presented in RZK (1997) where they assume similar angular 
momentum losses due to mass leaving the system from the primary.
In addition we find that although a minimum threshold is necessary in order 
to produce cycles at all, as the magnitude of $\eta$ increases, the mass range 
in which oscillations occur reduces. 

In the next section we discuss the effects of allowing
simultaneously two forms of consequential 
angular momentum losses: angular momentum losses
due to mass loss from the primary (e.g. nova explosions), and  
fluctuations of $\dot J$ coupled to irradiation--driven cycles. 
These results indicate that 
angular momentum loss due to mass loss from the primary or 
coupled CAML cycles, can extend
the irradiation--driven instability to lower periods and shift
its onset to smaller $\eta$ values.

\section{COUPLED CONSEQUENTIAL ANGULAR MOMENTUM LOSSES}
\label{Caml}

Angular momentum losses due to mass leaving the primary has some effect 
in enhancing mass transfer cycles as shown in the previous section. However, 
since it is an order of magnitude smaller than magnetic braking it is incapable 
of extending the mass transfer cycle very much. It is possible to get 
losses of the form (\ref{Irra.5}) to drive cycles 
at all orbital periods by assuming a much weaker magnetic braking, 
this however will cause the
size of the period gap to shrink to unacceptably small values. Thus
there is a need for another mass transfer dependent angular momentum loss in 
the system. 

As suggested in King et al (1996) we introduce an additional angular momentum 
loss mechanism which is proportionate to the instantaneous mass transfer rate  and
is thus a type of CAML.  Enhanced magnetic braking or some other mass transfer 
coupled angular momentum loss could be responsible for these fluctuations. We
show that from a combination of irradiation and 
$\dot J_{\rm cyc} / J\propto \dot M_2$, mass 
transfer cycles can be extended to all masses.

In previous work King et al (1996) already noted that the main reason the
irradiation--driven instability dies, as the secondary mass is reduced in
the course of mass transfer, is that the thermal timescale of the convective
envelope becomes too long. They also pointed out that 
oscillations in angular momentum losses directly coupled to the mass transfer 
cycle could extend the occurrence of limit cycles to lower companion masses.
They proposed a simple form of coupling which could be shown analytically 
to produce always an instability if the binary as a whole was pushed closer to 
dynamical instability. However, their analysis was limited to the onset of
the instability without investigating the form of the cycles 
nor their evolutionary effects. 
To determine if the resultant
cyclic evolution can be applied to CVs or other systems, detailed simulations
of the such cycles are necessary, allowing us to study its amplitude and 
time--dependence, and sensitivity to input parameters. 

We follow King et al (1996) and allow formally for coupled
variations in $\dot J$ writing the variations in the form
\begin{equation}
-{\dot J\over J} = {j(R_2, \dot M_2)\over t_J}\; , 
\label{Caml.1}
\end{equation}
where $t_J$ is the secular mean angular momentum 
loss timescale, and $j$ is a dimensionless function. 
To investigate the main effects of allowing for variations of the angular momentum
losses coupled to the transfer rate, King et al (1996)
considered the simplest possible case, in which these increase
linearly with the instantaneous $\dot M_2$, i.e.
\begin{equation}
j=1+l{\dot M_2\over [\dot M_2]_{\rm ad}},\quad l>0\; ,
\label{Caml.2}
\end{equation}
where $[\dot M_2]_{\rm ad}= -M_2/(D_s t_J)$ is the adiabatic 
transfer rate, and $D_s = (\zeta_s-\zR)/2$ is the standard
adiabatic stability denominator (See Section \ref{Stab}).
In equation (\ref{Caml.2}) above, the first term represents systemic losses while
the second term gives the fluctuating angular momentum losses coupled to the cycle 
$\dot J_{\rm cyc} / J = l \dot M_2/( [\dot M_2]_{\rm ad} t_J)$.
Formally, this is an example of a consequential angular momentum
loss (CAML, cf Webbink 1985; King \& Kolb, 1995): 
$\dot J/J$ is the sum of a $\dot M_2$--independent
``systemic'' component, represented by $1/t_J$, and the
``consequential'' components 
\begin{equation}
\frac{\dot J_{\rm CAML}}{J} ={\dot J_{\rm cyc} \over J}+  {\dot J_{\rm n.e.} \over J}\\
=l \; D_s \; \frac{\dot M_2}{M_2} +
\beta_1 M_2^2/(M M_1)\dMM \; .
\label{Caml.3}
\end{equation}
The stability of CAML is discussed in King \& Kolb (1995) in general terms.
King et al (1996) show that stability against dynamical--timescale
mass transfer requires $D=D_s(1-l) > 0$, i.e.\ $l<1$ when $\beta_1=0$. The 
case with arbitrary $\beta$ is discussed in Section \ref{Stab} below.

A sequence of evolutions with different values of the irradiation efficiency
$\eta$ is shown in Fig. 4, where we adopted $l=0.9$. In order to isolate the
effects of the CAML oscillations postulated by King et al (1996), we excluded
angular momentum losses due to nova explosions by taking $\beta_1=0$ in this example.
Cycles appear both above and below the gap but there is an intermediate range of masses
or correspondingly orbital periods in which mass transfer is stable. 
Wherever cycles occur, the mass transferred per cycle increases 
with $\eta$ but the total mass range over which mass transfer is unstable diminishes.
The combined effects of CAML from the primary and coupled oscillations
are illustrated by a sequence of models with $l=0.25$ and $\beta_1=1.0$ shown in
Fig. 5. Note that a reduction in $l$ is necessary to get qualitatively the same 
results as in a case with $\beta_1=0$ since both these effects increase the
coupling between CAML and mass transfer cycles. There are however some significant
differences between angular momentum losses due to nova explosions 
($\dot J_{\rm n.e.}/J$)  and cyclic angular momentum losses ($\dot J_{\rm cyc} / J$) 
as parameterized by $l$. 
Losses through novae have a more dominant effect on
higher mass systems than lower mass systems since they are $\propto {M_2}^2 /M$ for
a constant $\dot M_2/ M_2$. In contrast $\dot J_{\rm cyc}/J$ remains relatively 
constant with a constant $\dot M_2/ M_2$.  $\dot J_{\rm cyc}/J$ will decrease
some due to a shrinking $\zeta_s$, as the secondary becomes more 
convective, but will eventually increase as $\zR$ becomes more negative. Therefore
 $\dot J_{\rm n.e.}/J$ makes higher mass systems more unstable but has a
diminishing  effect as the secondary mass reduces. Since pure irradiation driven
mass transfer 
cycles are easier to produce at higher masses, $\beta_1$ does
not contribute much to extending the duration of cycles and only amplifies
the initial peak mass transfer rates. 
Another case with $l=0.95$, $\beta_1=0$ and $\eta=0.1$, is shown in Fig. 6, 
now plotted as a 
function of the binary orbital period. An example of CV evolution calculated using 
the G(x) tables obtained by Hameury \& Ritter (1997) from full stellar models
is shown on Fig. 7.  These examples 
demonstrate that a suitable adjustment of $l$ or the combined effects of 
$l$ and $\beta_1$ can produce cycles at all orbital periods. However,
the peak mass transfer rates are uncomfortably high and the period gap
is wider than in the other cases. This is a natural corollary of the fact that
CAML given by
equation (\ref{Caml.3}) can only enhance angular momentum losses form the
system and drive transfer rates {\em higher} than secular.

The above discussion suggests that the coupling
introduced in equation (\ref{Caml.2}) could be modified to read
\begin{equation}
j=1+l{(\dot M_2-[\dot M_2]_{\rm sec})\over [\dot M_2]_{\rm ad}},\quad l>0\; ,
\label{Caml.4}
\end{equation}
where $[\dot M_2]_{\rm sec}$ is the secular equilibrium rate under 
irradiation (King et al 1996). With this modification, there is no 
enhanced CAML if the system transfer mass at the secular rate and if
the transfer is stable. However, if there is an irradiation--driven
limit cycle, then coupled CAML occurs, and the net effect is to
enhance the cycle without changing the mass transfer averaged  
over secular timescales. An example of evolutions obtained with
this form of coupling is shown on Fig. 8. The peak mass transfer 
during cycles and the width of the period gap are now in line 
with observational limits. The same evolution calculated with 
an atmospheric scale height ten times smaller $\epsilon= 10^{-4}$
is shown for comparison in Fig. 9. The peak mass transfers in this case
are no more than a factor of 2 higher than in the calculation 
with $\epsilon= 10^{-3}$ at the same orbital period. The  mass and period 
changes during individual cycles are smaller as expected, but the overall 
evolution remains qualitatively unchanged.

\section{ANATOMY OF A CYCLE}
\label{Aml}
In order understand physically the nature of the cycles, it is interesting to 
analyze what happens during a typical cycle in a purely irradiation driven case 
and in a case where CAML is present, and to compare the results. As we shall see
below, pure irradiation cycles are driven by the thermal {\em expansion} of the 
secondary, while in CAML--assisted cycles at higher secondary masses one is 
enhancing angular momentum losses to the extent that it is the {\em contraction} 
of the Roche lobe which dominates. As the companion mass decreases, the thermal
expansion of the companion again dominates, but the net result is that 
irradiation--driven cycles now occur at shorter orbital periods.

Let us examine first the effects
of changes in the radius of the secondary and how it feeds back into changes
in the systemic orbital angular momentum losses.  
The systemic orbital angular momentum losses $\dot J_{\rm sys}$ 
respond to changes in $R_2$ since the magnetic braking angular 
momentum losses $\dot J_{\rm MB}/J\propto M_2R_2^4/a^5$. 
We refer to Fig. 10 which shows results obtained for a pure
irradiation--driven cycle to illustrate the discussion given below.
Examining the bottom panel of Fig. 10, we see that the rate of change of
magnetic braking goes through six distinct stages corresponding to
distinct behavior identified by numeric labels. 
These stages are correlated to rates of change in 
$R_2$, $M_2$ and $a$ (shown in the middle panel Fig. 10).
Initially, before mass transfer starts,
as the star approaches Roche Lobe contact, the companion
is near equilibrium and its mass and radius are virtually constant.
Hence $\dot J_{\rm MB}/J\propto a^{-5}$, increasing 
rapidly as $a$ decreases (stage 1). Once contact is reached the secondary 
begins to expand due to irradiation, $a$ is still decreasing, thus 
driving an even stronger angular momentum loss (stage 2). As the mass transfer 
rate increases, $\dot a$ changes sign and the orbital separation begins 
expanding at an increasing 
rate. Even though the radius of the secondary is still expanding during 
this phase, the combination of enhanced mass transfer and orbital expansion 
rapidly reduces the braking (stage 3).  After the peak rate of 
orbital separation and mass transfer is reached, the rate of decrease in 
braking slows down as the secondary is still expanding but this stage is
relatively short lived (stage 4).  At 
some point the effects of irradiation saturate and the companion
starts to shrink towards its new equilibrium radius. Thus, initially,
the magnetic braking rate decreases
as  $\dot J_{\rm MB}/J \propto R^{4}$, (stage 5) since on this time scale both $a$ 
and $M_2$ are constant. As the secondary adjusts thermally, the magnetic braking 
rate returns (stage 6) again to the detached behavior 
$\dot J_{\rm MB}/J \propto a^{-5}$ (stage 1). 
During the nearly detached phase the system is slowly driven back into
contact and the cycle repeats.  
 
We also present results for two individual cycles selected from the run with 
CAML--assisted cycles shown in its entirety on Fig. 9.  The cycle shown
on Fig. 11 occurs early in the evolution when $M_2\approx 0.9 \msun$. 
This evolution was calculated with $\epsilon=10^{-4}$ and therefore, during 
contact, $R_2$ and $\RL$ follow each other more closely than in the case shown 
in Fig. 10. It is clear from the top panel of Fig. 11 that the enhanced braking 
during high mass transfer leads to a contraction of the binary, the Roche lobe 
and the companion.  In the early stages of contact, as the mass transfer increases,
the secondary expands in response to irradiation but eventually the effect of
CAML dominates the evolution and results in a rapid overall contraction.
As the companion is trying simultaneously to 
adjust thermally to the incident irradiation, a high peak of mass transfer results.
Later, in the same evolution, once the companion's mass has decreased to
$M_2\approx 0.4\, \msun$, the thermal expansion of the secondary dominates
during a typical cycle as shown in Fig. 12. The reasons for this behavior   
are twofold: 1) the specific angular momentum of the primary is reduced, and
2) the mass transfer rate is lower. As a consequence both CAML terms in 
equation (\ref{Caml.3}) are smaller than before and the mechanical tendency
for the system to expand during high mass transfer episodes wins. These 
effects translate in accompanying period variations which are discussed in
Section \ref{Pvar} below.

\section{PERIOD VARIATIONS}
\label{Pvar}
The balance between the the systemic angular momentum losses and the 
mechanical tendency for the orbit to expand during mass transfer yields 
in the standard evolution a stable mass transfer rate which changes only 
secularly. However, in evolution under irradiation, this balance is 
temporarily affected by the cycles resulting in orbital period changes of 
either sign. 
Generally one expects that during the bright phases of the cycle
the orbit expands on timescales $\sim$ few $\times 10^6 - 10^7$ yrs  
as the companion expands. On the other hand, during the faint 
quasi--detached phase of
the cycle the system contracts undetected and the overall effect is
a secular evolution toward shorter periods. The simulations shown
on Fig. 13 illustrate
that this is indeed the case for irradiation--driven cycles without 
CAML, which can only occur at longer orbital periods. 
This is potentially interesting because it would predict a systematic
effect, supposing one could disentangle these intermediate time scale
variations from other short term variations: bright systems should
statistically show a {\em large} and {\em positive} $\dot P$. 

However, when CAML is included the calculations show that enhanced
magnetic braking can actually dominate and cause orbital contraction at 
longer periods. In that case, if CAML remains at a sufficient 
level, cycles can occur at all orbital periods with the bright phase 
of the cycle yielding  $\dot P<0$ at long periods and  $\dot P>0$ just 
above and below the gap (see Fig. 14). This behavior is easily understood 
analytically if one considers the logarithmic derivative of the orbital 
period including all mass and angular momentum losses present,
\begin{equation}
	 {\dot P \over P} = {\dot J \over J}_{sys}+{\dot M_2 \over M_2} 
	 \left[ -3+(1-\alpha){q \over 1+ q}+3 \alpha q+ 3 \beta q 
	 \left( {q \over 1+q} \right) +3D_s l \right]\; ,
\label{Pvar.1}
\end{equation}  
where $\dot M_1 = -\alpha\dot M_2$ has been asssumed.    
The last two terms are the contributions due to losses associated with 
CAML (mass loss from the primary and coupled fluctuations). 
Without these terms the contribution
to $\dot P$ which is proportional to mass loss rate is always positive. Thus
in the pure irradiation case in which mass loss is enhanced, large positive
orbital derivatives occur during bright phases. In CAML--assisted cycles,
the increase in mass loss during irradiation--driven expansion of the secondary 
also increases $\dot J_{\rm CAML}$ to the point where angular momentum losses
dominate over the mechanical tendency of the system to expand upon mass exchange.
Eventually as $M_2$ decreases, the specific angular momentum carried away by
mass loss decreases and the expansion tendency during high mass transfer,
dominates. The expansion becomes more pronounced as $M_2$ is reduced
occurring at smaller peak mass transfer rates and for longer durations.

The above suggests that further studies may lead to a
statistical observational test of the picture. If there was a certain
period range over which the period variations during bright phases had a 
definite sign,  one could in principle deduce the relative importance 
of thermal and CAML effects, for example.  It is, however, 
important to note that period variations occur in a variety of time scales.
Over the longest, secular time scale $\gta 10^8$ yrs, the period should
decrease as long as the companion is non--degenerate.  The period
variations induced by cycles occur at intermediate time scales, 
$P/\dot P \gta 10^6$ yrs and should be easier to observe than the secular trend.
However, other ``short" term effects have been observed, presumably caused
by stellar processes such as magnetic cycles, starspots etc., with time scales 
($P/\dot P \gta 10-100 $ yrs). Just as cycles can dominate the secular
trend, these short term effects may make it difficult to detect changes due to
thermal and CAML cycles. Nevertheless, it is worth pursuing this test
since one could learn much about processes governing the evolution of 
close binaries if shorter term orbital effects could be successfully
filtered out.

\section{STABILITY CONSIDERATIONS}
\label{Stab}
     In order to analyze the stability of mass transfer in semi--detached
compact binaries under the 
influence of irradiation, mass loss from the primary and CAML cycles,
 it is necessary to examine how the radius of the mass
donor star, $R_2$ changes in comparison to the size of the Roche lobe, 
$R_{\rm L}$.
When  $R_2$ expands faster than $R_{\rm L}$ we get unstable mass transfer,
 whereas if $R_2$
expands slightly slower than $R_{\rm L}$, mass transfer can greatly decrease 
or even stop altogether. It is only in the case were $R_2$ and $R_{\rm L}$ 
expand at the same rate that stable mass transfer occurs.
     Changes in $R_2$ can be expressed as (Ritter 1995):
	  \begin{equation}
	  {\dot R_2 \over R_2} = \zeta_s
	      {\dot M_2 \over M_2} + \left({\partial \ln{R_2} 
	      \over \partial t} \right)_{\rm nuc}+ \left( {\partial \ln{R_2}    
	      \over \partial
	      t}\right)_{\rm th} - \zeta_{\rm irr}{\dot M_2 
	      \over M_2} \; ,
 \label{Stab.1}
 \end{equation}
where the adiabatic response to mass loss is described by the
adiabatic mass--radius exponent 
\begin{equation}
	     \zeta_s=\left( {\partial \ln{R_2} \over \partial 
	     \ln {M_2}}\right)_{\rm ad} \; ,
 \label{Stab.2}
 \end{equation}
the second and third terms describe the effects of nuclear evolution and
thermal relaxation respectively, 
and the rate of expansion due to irradiation has been written in 
	analogy with the adiabatic expansion, using an 
	``irradiation mass--radius exponent", defined by
 \begin{equation}
   \zeta_{\rm irr}=- M_2 {\partial \over \partial \dot M_2}\left( {\partial
	    \ln{R_2} \over \partial t} \right)_{\rm th} \; .
 \label{Stab.3}
 \end{equation}
	 
Similarly, changes in the Roche-Lobe radius can be expressed as:
	 \begin{equation}
	   {\dot R_{\rm L} \over R_{\rm L}} = \zR {\dot M_2 
	   \over M_2}+2\left({\partial
	    \ln{J} \over \partial t}\right)_{\dot M_2=0}
	    -\zeta_{\rm CAML}{\dot M_2 \over M_2}
 \label{Stab.31}
 \end{equation}
  where the term with $\zR$ describes changes due to redistribution of 
mass and angular momentum in the system, $(\dot J/J)_{\dot M_2=0}$ 
is the rate of systemic angular momentum losses, and the term containing
 \begin{equation}
	    \zeta_{\rm CAML}=-2 M_2 {\partial \over \partial \dot M_2}
	    {\partial
	    \ln{J} \over \partial t}
 \label{Stab.4}
 \end{equation}	
gives the rate of change due to CAML. 
	   By equating (\ref{Stab.1}) with (\ref{Stab.31}) and solving for 
	   $\dot M_2$   for stationary mass transfer, we get:
 \begin{equation}
	   -\dot M_{\rm 2st}={M_2 \over \zeta_s-\zR-
	   \zeta_{\rm irr}+\zeta_{\rm CAML}}
	      \times \left[\left({\partial \ln{R_2} 
	      \over \partial t}\right)_{\rm nuc}+
	      \left({\partial \ln{R_2} \over \partial
	      t}\right)_{\rm th}-2\left({\partial \ln{J} \over \partial t } 
	      \right)_{\dot M_2=0} \right] 
\label{Stab.5}
\end{equation}   
	   which leads  to the criterion for adiabatic stability:
 \begin{equation}
	   D_{\rm ad}=\zeta_s-\zR-\zeta_{\rm irr}+
	   \zeta_{\rm CAML}>0
\label{Stab.6}
\end{equation}
For a system driven only by systemic angular momentum losses ,
in the absence of irradiation and CAML, equation (\ref{Stab.6}) yields 
the standard result $-\dot M_{\rm 2st}= D_s^{-1}(\dot J/J)_{\dot M_2=0}$,
which is adiabatically stable when the standard adiabatic denominator
$D_s = (\zeta_s-\zR)/2$ is positive.
If we assume that  consequential angular momentum loss (CAML) is 
proportional to the instantaneous mass loss rate and angular momentum 
losses due to 
mass leaving the primary can be parameterized by $\beta_1$ we can 
write these angular momentum losses as in equation (\ref{Caml.3}):
\begin{equation}         
	   \biggl({\dot J \over J}\biggr)_{\rm CAML}=  D_s l\dMM+ 
	   \beta_1 M_2^2/(M M_1)\dMM \; ,
\label{Stab.6.1}
\end{equation}
 which using the definition (\ref{Stab.4}) gives 
	   \begin{equation}         
	   \zeta_{\rm CAML}= -2 D_s l-2 \beta_1 M_2^2/(M M_1)\; .
\label{Stab.7}
\end{equation}
 Adopting Paczy\'nski's approximation for the Roche Lobe radius, $R_{\rm L}/a=f(q)$:
         \begin{equation}
         {R_{\rm L} \over a}= {2 \over 3^{4 \over 3}} \left({q \over 1+q}\right)^{1 \over 3} = .462 \left({M_2 \over M}\right)^{1 \over 3}, q \lta 0.8
            \label{e2.0a1}
            \end{equation}  
 and 
the definitions $q=M_2/M_1$ and $\dot M_1= -\alpha \dot M_2$, a general
expression for $\zR$ is obtained
\begin{equation}
	 \zR =  -{5 \over 3}+{2 \over 3}(1-\alpha){q \over 1+ q}+2 \alpha q \; .
\label{Stab.8}
\end{equation} 
Therefore, for conservative mass transfer ($\alpha=1$, no mass is lost from the binary)
\begin{equation}         
	   \zR = -5/3 +2 M_2/M_1\; ,
\label{Stab.71}
\end{equation}
whereas for non-conservative mass transfer ($\alpha=0$, all matter transferred is eventually lost)
\begin{equation}
	 \zR = -5/3 +2 M_2/ 3M \; .
 \label{Stab.72}
\end{equation}
The use of Paczy\'nski's approximate formula rather than the more exact 
Eggleton's formula (see e.g. Frank, King \& Raine 1992)
for stability considerations will give approximate 
stability boundaries, but is convenient for simplicity. Furthermore,
for the mass range that we are considering, this approximation 
is accurate enough for most numerical work.

During the cyclic mass transfer phases mass transfer has become 
adiabatically unstable due to the irradiation expansion term of $R_2$.
The bipolytrope code provides an adiabatic expansion term which  
for higher masses agrees in sign with
the full stellar model but is somewhat smaller in magnitude. This results in 
stars with mass $\gta 0.7\, \msun$ being slightly more unstable than a full
stellar model would predict. The 
agreement for stars with mass $\lta 0.7\, \msun$ is good and improves as
$M_2$ decreases.
Now that we have expressions for $\zR$ , $\zeta_{\rm CAML}$, 
and $\zeta_s$ is obtained from our bipolytropic code, an explicit expression
for $\zeta_{\rm irr}$ is all we need to apply the stability criterion
given by $(\ref{Stab.6})$.  From (\ref{Stab.3}) we have:
\begin {equation}
\zeta_{\rm irr}= - M_2 {\partial \over \partial L} \left({\partial 
\ln{R_2} \over \partial t} \right) {dL \over d \dot M_2}
\label{Stab.725}
\end{equation}
 Computing $(\partial^2 \ln{R_2} / \partial t \partial L)$ 
 in (\ref{Stab.725}) from the bipolytrope model (e.g. KR92) we obtain
\begin{equation}
{\partial \over \partial L}\left( {\partial
	    \ln{R_2} \over \partial t} \right)_{\rm th}=-{R_2 
	    \over GM_2^2} {\cal F}(Q, n_1) 
\label{Stab.73}
\end{equation}
where ${\cal F}(Q, n_1)$ is a numerical function calculated by the 
bipolytropic code.
This leads to a generalized stability criterion similar to the 
$\Lambda$ criterion given by RZK [eq. (59)] and to equation (5) 
from Hameury and Ritter.
\begin {equation}
    \Lambda=2(\zeta_s-\zR+\zeta_{\rm CAML}) 
    {\tau_{\rm KH} \over \tau_{\dot M_2}} {\cal F}^{-1}(Q,n_1) > {1 \over 2 \pi} 
\int_0^{\theta_{\rm max}} x(\theta)g[x(\theta)] \sin{\theta} d\theta= I_{\rm ps}
\label{Stab.74}
\end{equation}
Where, following notation introduced in RZK,  we define $g[x(\theta)]= -{dG / dx}$, 
$\tau_{\dot M_2}=M_2 / {\dot M_2}$  
the mass transfer timescale and $\tau_{\rm KH}$ is the Kelvin-Helmholtz timescale.
 In RZK the stability criterion is derived from one of several
starting points:  any one of their equations (25), (40) or
(52) can be used, taking into account the definition of $\Lambda$, but the
result does 
not include the effects of any CAML.  Angular momentum losses 
through the primary 
are incorporated into the definition of $\zR$ in RZK whereas we
include the effect explicitly in the definition of $\zeta_{\rm CAML}$.
In RZK, the stability criterion -- in the absence of CAML --
can be written $\zeta_s-\zR > \zeta_{\rm irr}$ where $\zeta_{\rm irr}$ can 
be expressed after some work as proportional to the r.h.s. of 
(5) in Hameury and Ritter (1997).  The criterion  is written in this way in order to 
group in $\Lambda$ terms that are related to the stellar and binary structure on 
the left hand side, while the 
terms on the right hand side depend on the modeling of the irradiation.
 
We present two examples to illustrate the application of the stability criterion. 
In each case we plot  $\Lambda$ and $I_{\rm ps}$  versus $M_2$ calculated 
along a fictitious evolution in which irradiation is present and affects the
stationary mass transfer, but cycles are suppressed.  This is necessary since the
stability criterion is based on a {\em linear} analysis of small departures from the 
stationary mass transfer (the ``fixed points" of the
equations; see King et al 1996 for details). In the regions in which $I_{\rm ps}$ is 
greater than $\Lambda$ we anticipate the 
system will be unstable and experience 
mass transfer cycles.  The bottom panel in each graph shows the complete
{\em non-linear} evolution with cycles generally occurring where 
$I_{\rm ps}>\Lambda$. The case shown on Fig. 15 starts unstable, the instability weakens 
around $0.7\, \msun$, but strengthens around  $0.6\, \msun$, and finally 
stabilizes above the gap. The mass transfer below the gap is predicted to be stable but 
a few oscillations occur before the mass transfer settles down to its stationary value. The 
case shown on Fig. 16 is unstable throughout according to the top panel
and the non-linear simulation shown below confirms this. Therefore the
linear criterion works quite well but the remaining 
differences indicate that ultimately a non-linear simulation is necessary since 
large departures from the stationary points occur during cycles.
  
The stability criterion (\ref{Stab.6}) shows that any form of CAML with
$\zeta_{\rm CAML}<0$ reduces the adiabatic denominator 
$D_{\rm ad}$ and makes mass transfer more unstable. 
The simulations presented in this paper do indeed show that $\beta_1$ and 
$l$ nonzero and positive make mass transfer more unstable, as expected from
equation (\ref{Stab.7}), thus extending the range 
of the irradiation--driven cycles .

\section{DISCUSSION AND CONCLUSIONS}
\label{Conc}
The evolutionary calculations presented in this paper incorporate the 
effects of irradiation of the companion and of consequential angular 
momentum losses (CAML). These simulations confirm previous analytic and 
numerical work on pure irradiation--driven cycles, explore secular evolution
with CAML--assisted cycles, examine in detail what happens during a cycle,
analyze orbital period variations, and extend the discussion of stability to
the more general case including CAML.
 
Our results show that mass transfer cycles can occur at all orbital periods
when CAML is strong enough. Therefore irradiation-driven
cycles assisted by CAML may account for the dispersion observed in estimates of
the mass transfer for CVs at a given orbital period without blurring significantly
the period gap. The main weakness of the model in its present form is the absence
of a physical mechanism that would enable to predict the strength of the 
coupling between CAML fluctuations and mass transfer, and to ascertain its
dependence on system parameters. It is conceivable that the rate of
magnetic braking for example could be affected by irradiation, by varying 
the underlying wind mass loss rate or by altering the state of ionization
of the wind, but so far, no working model exists for these effects.
In the meantime one can assume a certain form for the coupling and 
follow the secular evolution, predicting the properties of the resulting
cycles and comparing these with observations.
For example, the correlations 
predicted between the sign of the orbital period derivative over intermediate 
time scales, the orbital period 
itself and the relative importance of CAML may serve as a test of the picture.

Another test may be provided by non-magnetic CVs 
with accretion disks 
if one adopts the
premise that their outburst properties are the combined result of 
the irradiation instability determining the accretion rate over intermediate 
time scales, and the disk instability determining the short term behavior
(see e.g. Cannizzo 1993; 1997 for reviews). Above a critical accretion
rate the disk will be hotter than the hydrogen ionization
temperature throughout, and thus stuck in the hot state (Smak 1983). 
Estimates for this critical rate depend on the assumed size and shape of the disk,
and on the effects of irradiation on the disk temperature and vertical structure
(van Paradijs 1996). In any case, the disk instability may be 
simply assumed not to occur above some critical
mass transfer rate $\dot M_{\rm crit}$ which is itself subject to
some uncertainty.  This approach is 
adequate if one is interested only in the overall outburst behavior (e.g.
presence/absence of dwarf nova outbursts) rather than the detailed shape of 
individual outbursts. 

A well--known example of the insight one may gain from studies of the 
period distribution is provided by the relative paucity of dwarf novae 
relative to nova--like variables at orbital periods just above the gap,
in the range $3-4$ hours (Shafter, Wheeler \& Cannizzo 1983; Shafter 1992).
The simulations presented here
suggest that irradiation--driven cycles assisted by a certain amount 
of coupled CAML could produce this effect in one of two ways:

1) if a moderate CAML is assumed, then mass transfer oscillations
will damp out at periods around 4 hours. Depending on the efficiency of the 
irradiation assumed (how effective is the blocking),  the mass transfer rate
between the periods of 3--4 hrs above the gap may be above the critical mass 
transfer rates for dwarf novae and be observed as novae like systems.

2) if a stronger CAML is assumed, cycles do occur at periods 3--4 hours,
but  most systems still spend most of the time 
they are visible, accreting at rates above the critical (maximum) rate for 
disk outbursts and thus appear as nova--like most of the time.

 An important caveat to this scenario is that 
the gap width is very sensitive to the mass loss timescale at the upper 
edge of the gap. The width of the period gap depends on the ratio of the mass 
loss timescale to the thermal timescale of the secondary star. This led us 
to explore a coupling (\ref{Caml.4}) different in form from the original suggestion 
of King et al (1996) which yields relatively lower mass transfer rates at the 
upper edge of the gap, and yet produces cycles at all orbital periods.
A population synthesis study incorporating the cyclic evolution described here
will be carried out during future investigations in order to determine if the 
predicted observational properties of
such an ensemble of CVs agree with existing observational limits. A particularly
interesting result would be a distribution of CV subtypes over the observed 
orbital period range. Such a 
study should enable us to place constraints on the possible mechanism(s) 
coupling CAML and irradiation--driven cycles. Given the uncertainties mentioned 
above we postpone a detailed modeling of this effect to a forthcoming paper. 

Finally, on a more speculative note, we note that in some of
the CAML--assisted evolutions, it is possible to get mass transfer 
rates $\sim$ few $10^{-7}\, \msun$/yr for $10^{5-6}$ yrs. Perhaps it 
is possible to get mass transfer rates in this range extending 
down to orbital periods of $3-4$ hr by adjusting
the CAML losses adequately. It is tempting to identify such
systems with the short period CV--type supersoft X--ray sources 
(e.g. RX J0439.8--6809 in the LMC and 1E0035.4--7230 in the SMC)
which do not conform to the generally accepted model in which the companion is
more massive than the white dwarf (Kahabka \& van den Heuvel 1997). 

We are grateful to Andrew King, Uli Kolb, Hans Ritter and Jean--Marie Hameury
for helpful discussions at various stages of this work. 
This work was partially supported by NSF grant AST-9020855 and 
NASA grant NAG 5-3082 to LSU. JF acknowledges the kind hospitality of 
the STScI where part of this work was done.

{}

\clearpage

\begin{figure}
\plotone{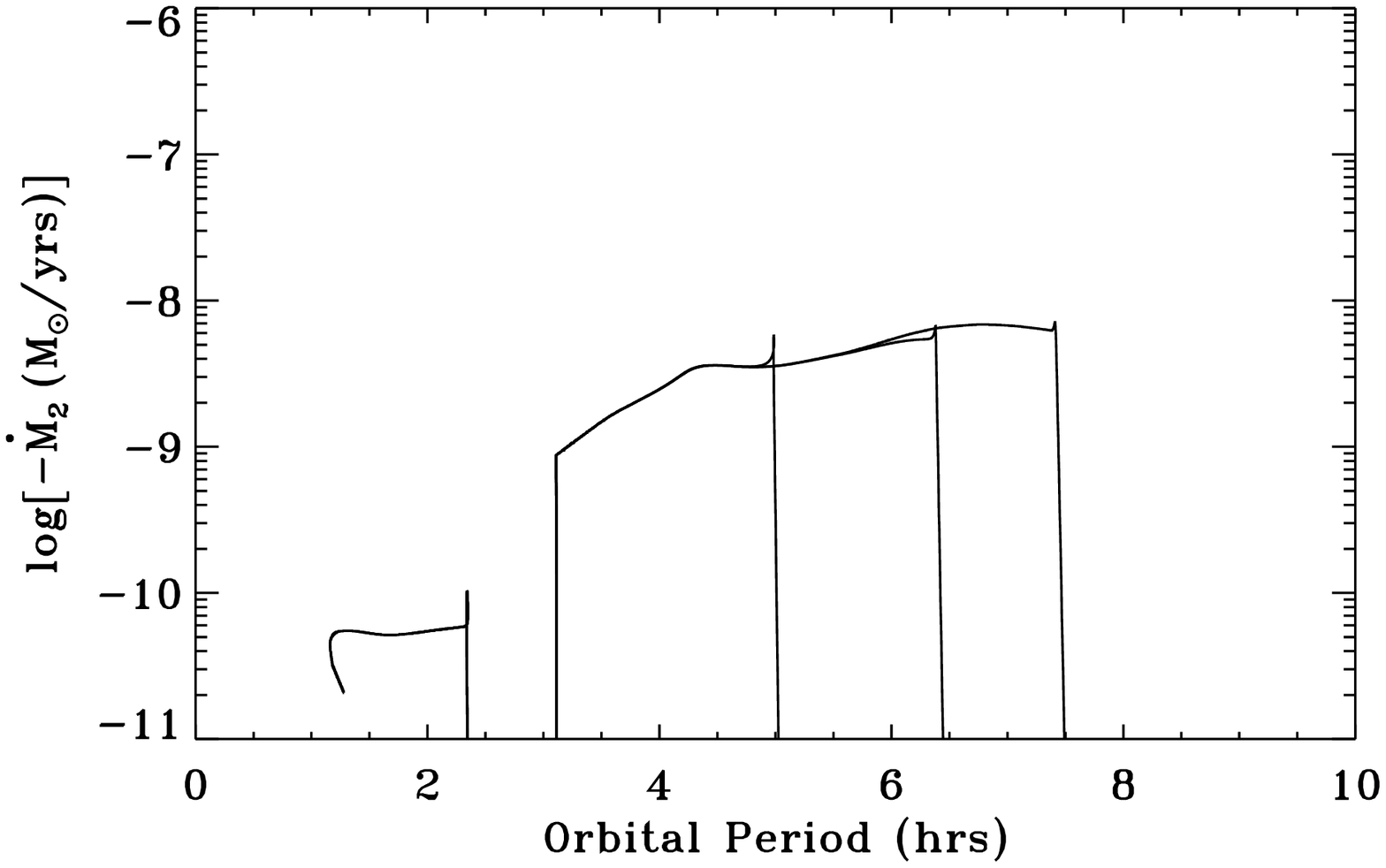}
\caption{Examples of standard CV evolutions without irradiation.
Three evolutionary paths with initial companion masses $M_2 =0.6,0.8,1.0\,\msun$
are shown to come into contact at different periods and converge
at shorter periods. The period minimum, the position and width of
the period gap are virtually the same.}
\end{figure}

\begin{figure}
\plotone{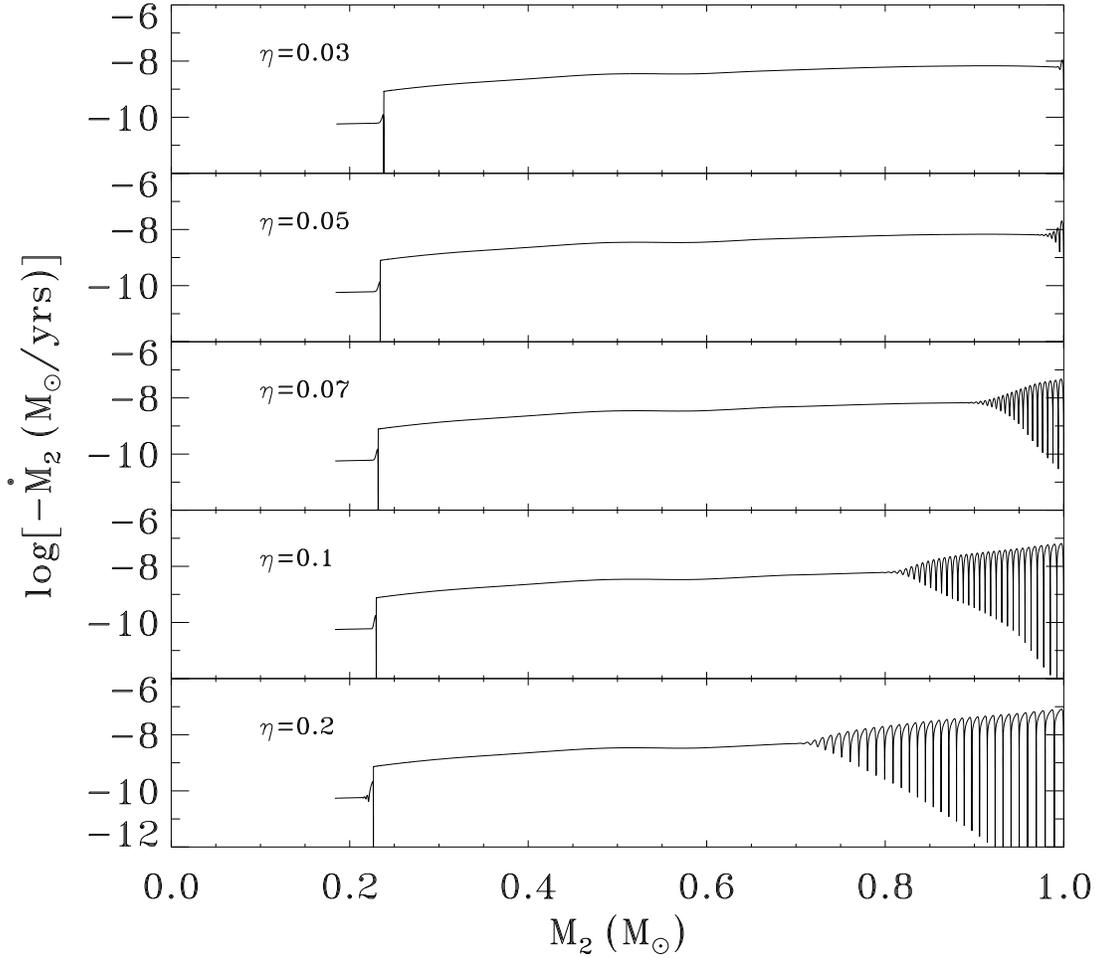}

\caption{A set of evolutions with increasing irradiation efficiency
$\eta$ values from top to bottom.
The mass transfer rate is shown here as a function of companion mass.
The discontinuity occurs when the companion becomes fully convective, the 
binary detaches and enters the period gap.
Limit cycles below the period gap occur in the last 
$\eta=0.2$ model. All the models have some mass transfer oscillations but 
they only
become significant with $\eta \gta 0.05$. It is assumed that $M_1=$ const. and
no angular momentum is lost with the mass leaving the system (e.g. in nova 
explosions).}
\end{figure}

\begin{figure}
\plotone{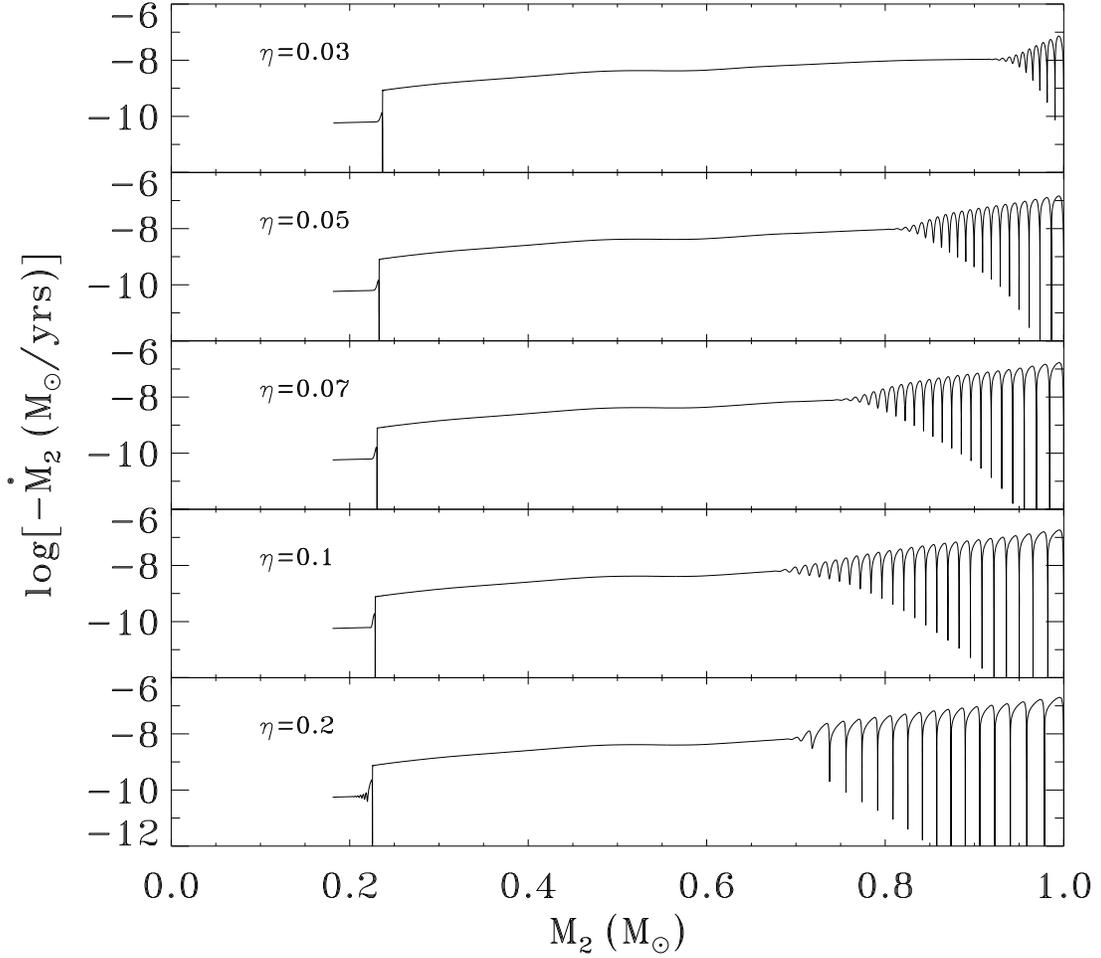}
\caption{Same as Fig. 2, but with mass loss carrying away the specific 
angular momentum of the primary, $\beta_1=1.0$ (see text for definitions).
Mass transfer oscillations below the period gap occur in the last 
$\eta=0.2$ model. All cases shown have some significant mass transfer 
oscillations.}
\end{figure}

\begin{figure}
\plotone{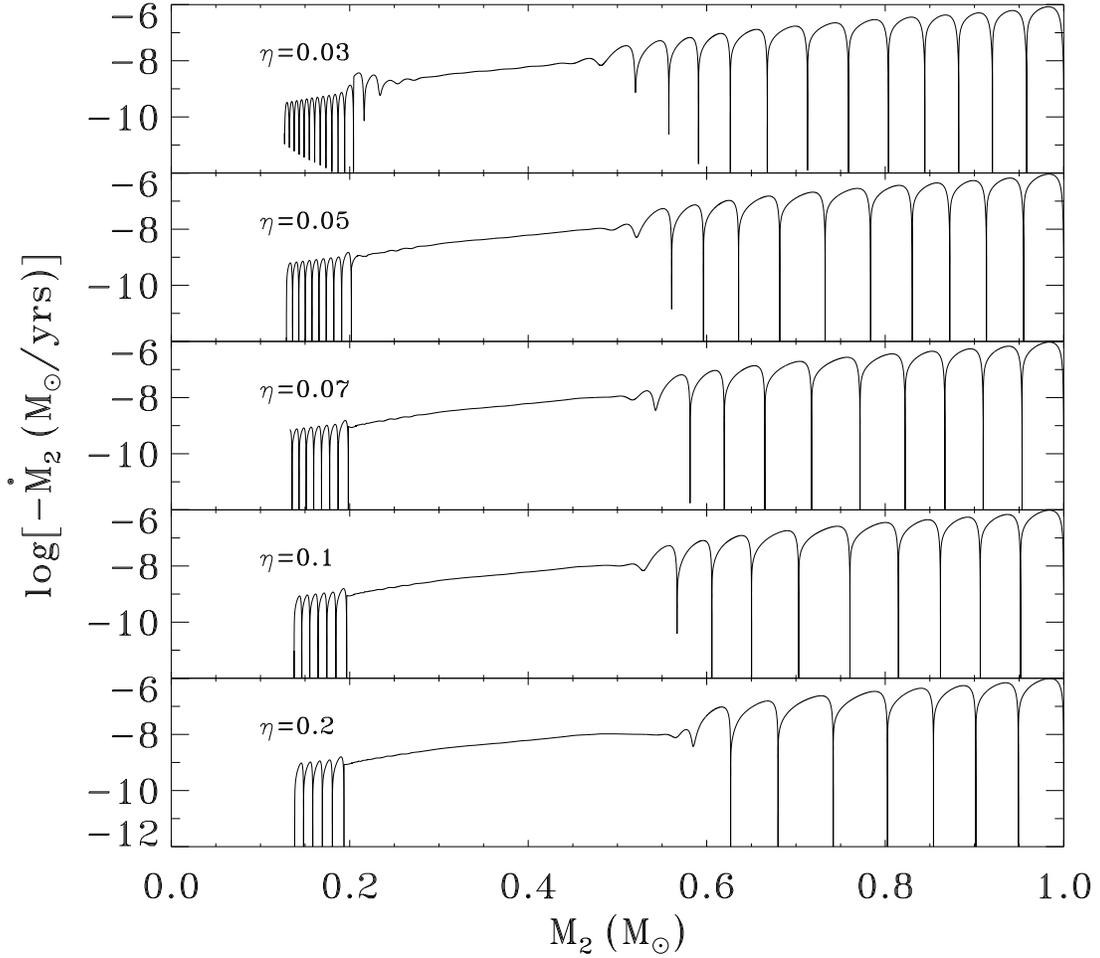}

\caption{Same as Figs. 2 and 3 with systemic angular momentum oscillations
coupled to the limit cycle with $l=0.9$. In order to isolate the effects
of such cycles, no CAML due to
mass loss from the primary $\beta_1=0.0$ was included in this case.
Mass transfer oscillations below the period gap occur all models. All the models have some 
significant mass transfer 
oscillations. The amplitude of oscillations increases with 
larger $\eta$ values but the overall duration decreases. }
\end{figure}

\begin{figure}
\plotone{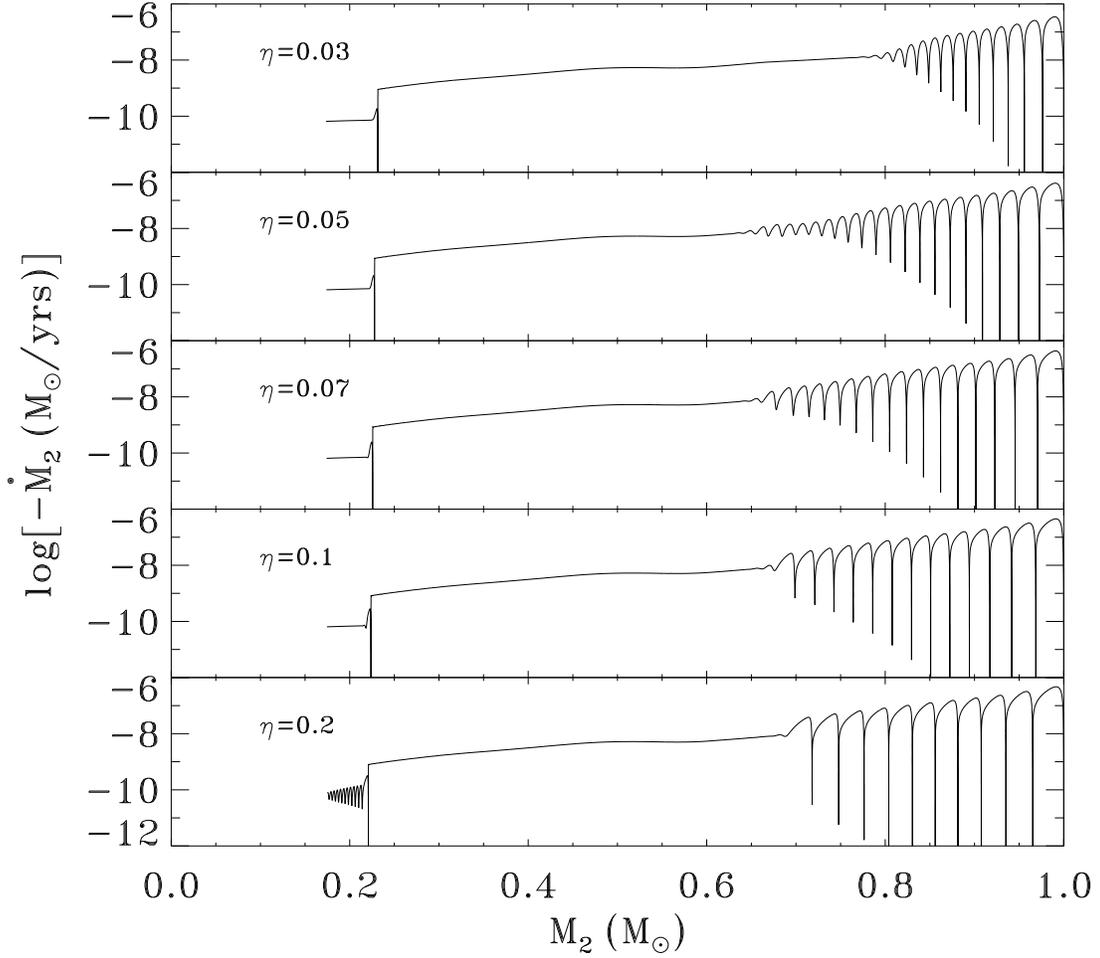}
\caption{Same as Figs. 2--4 with systemic angular momentum oscillations
coupled to the limit cycle with $l=0.25$. CAML due to
mass loss from the primary was assumed to leave the system with the specific angular 
momentum of the primary i.e. $\beta_1=1.0$.
Mass transfer oscillations below the period gap occur in the last 
$\eta=0.2$ model. All the models have some significant mass transfer 
oscillations. The amplitude of oscillations increases with 
larger $\eta$ values. The duration initially increases  with 
larger $\eta$ values but later decreases for larger $\eta$ values.}
\end{figure}

\begin{figure}
 \plotone{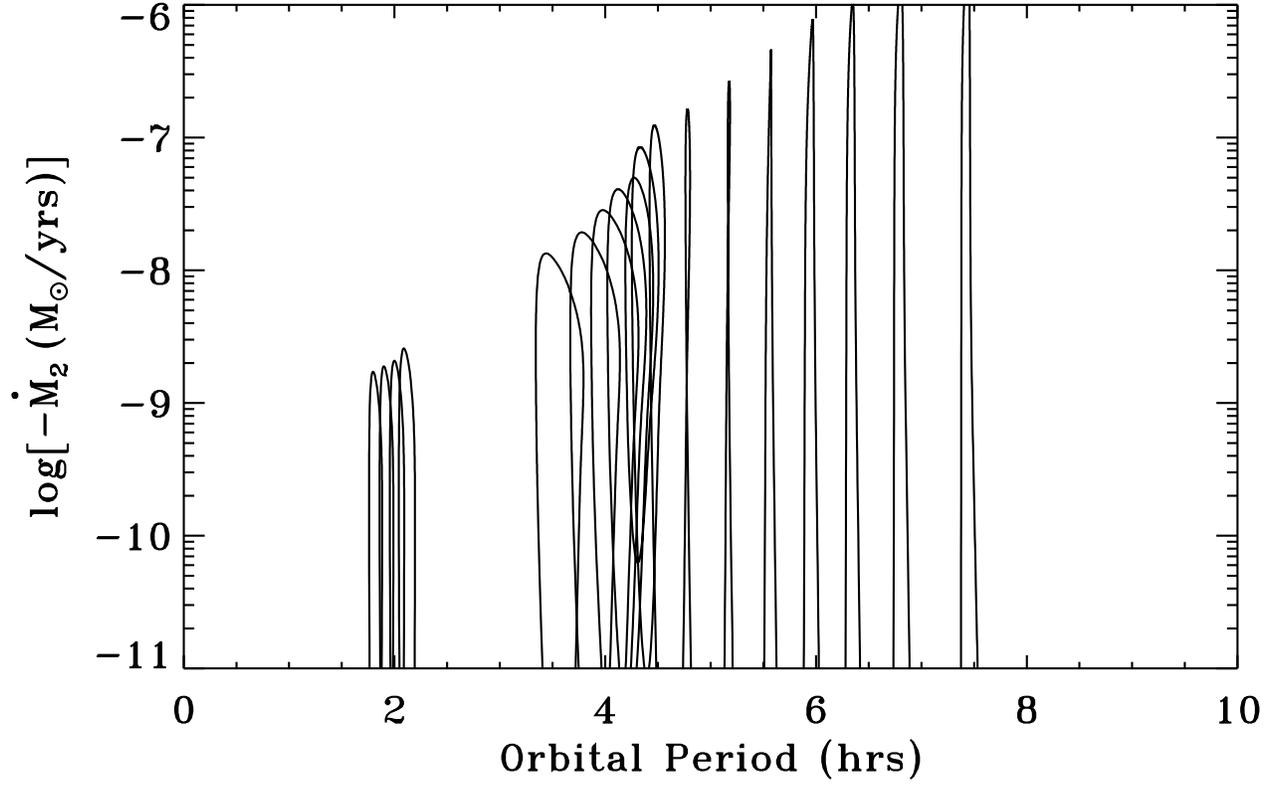}
\caption{A CV evolution with mass transfer cycles which extend to all orbital 
periods as a result of coupled CAML fluctuations in the form given by equation 
(\ref{Caml.3}), with $l=0.95$, $\beta_1=0$, and $\eta=0.1$.}
\end{figure}

\begin{figure}
\plotone{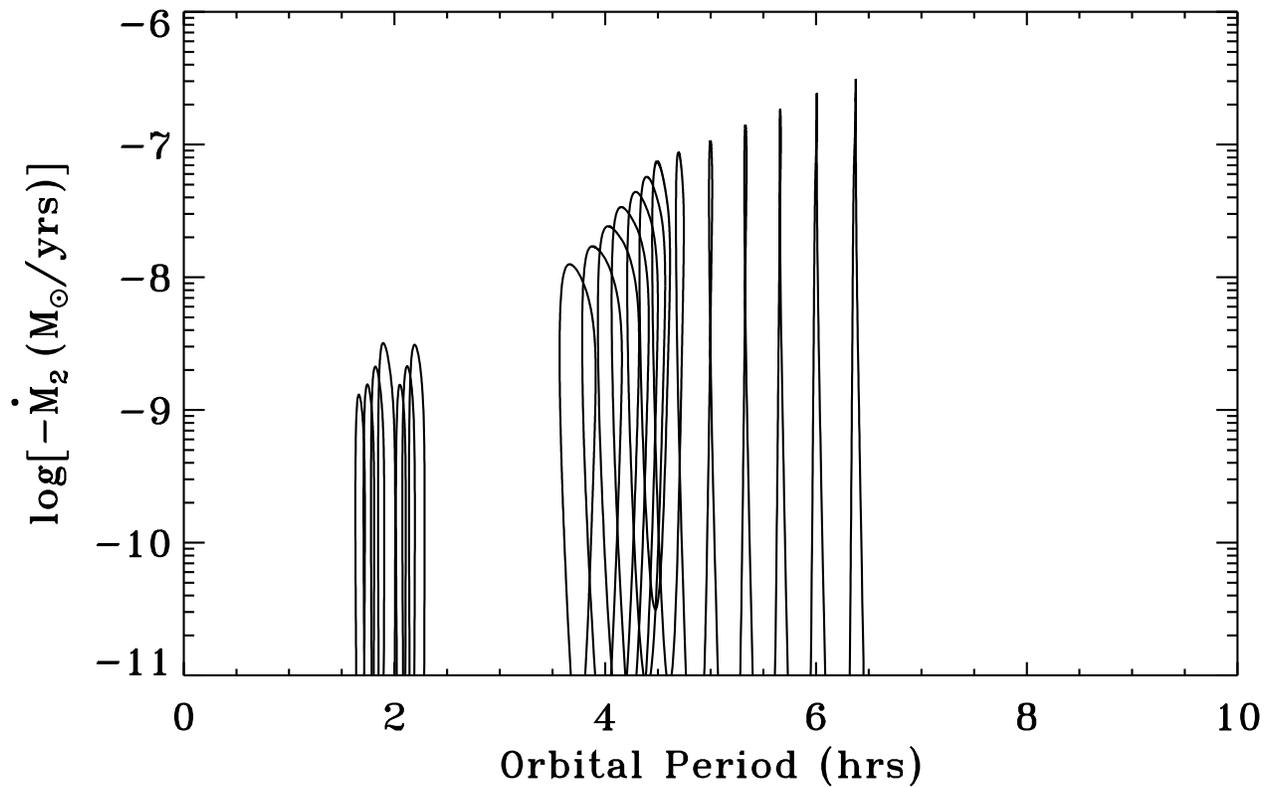}
\caption{An evolution with coupled CAML in the form given
by equation (\ref{Caml.3}), with $l=0.95$, $\beta_1=0$, and $\eta=0.01$, but 
making use of tabular solutions for G(x) for full stellar models
provided by Hameury and Ritter (1997).}
\end{figure}

\begin{figure}
\plotone{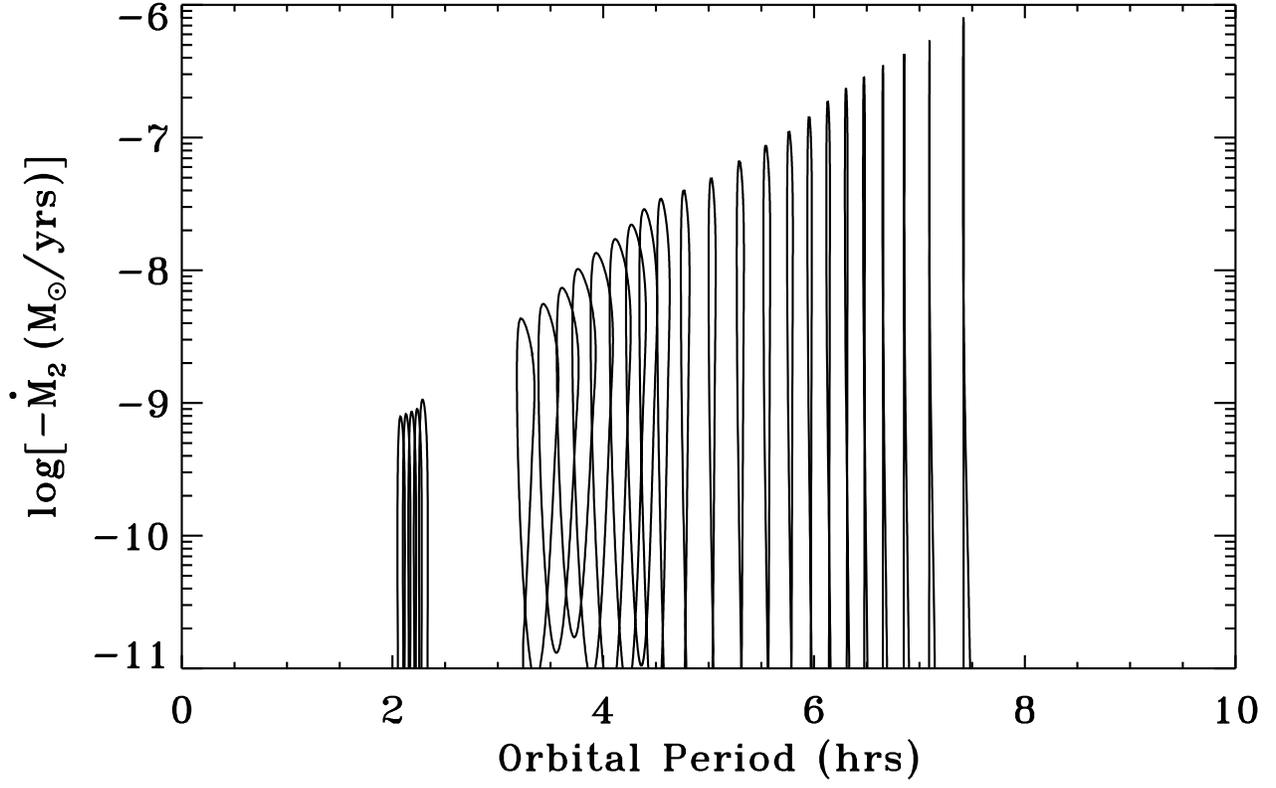}
\caption{A CV evolution with mass transfer cycles which extend to all periods 
calculated with CAML fluctuations in the form given by equation (\ref{Caml.4}), 
with $l=.9$, $\beta_1=0$ and $\eta=.05$.}
\end{figure}

\begin{figure}
\plotone{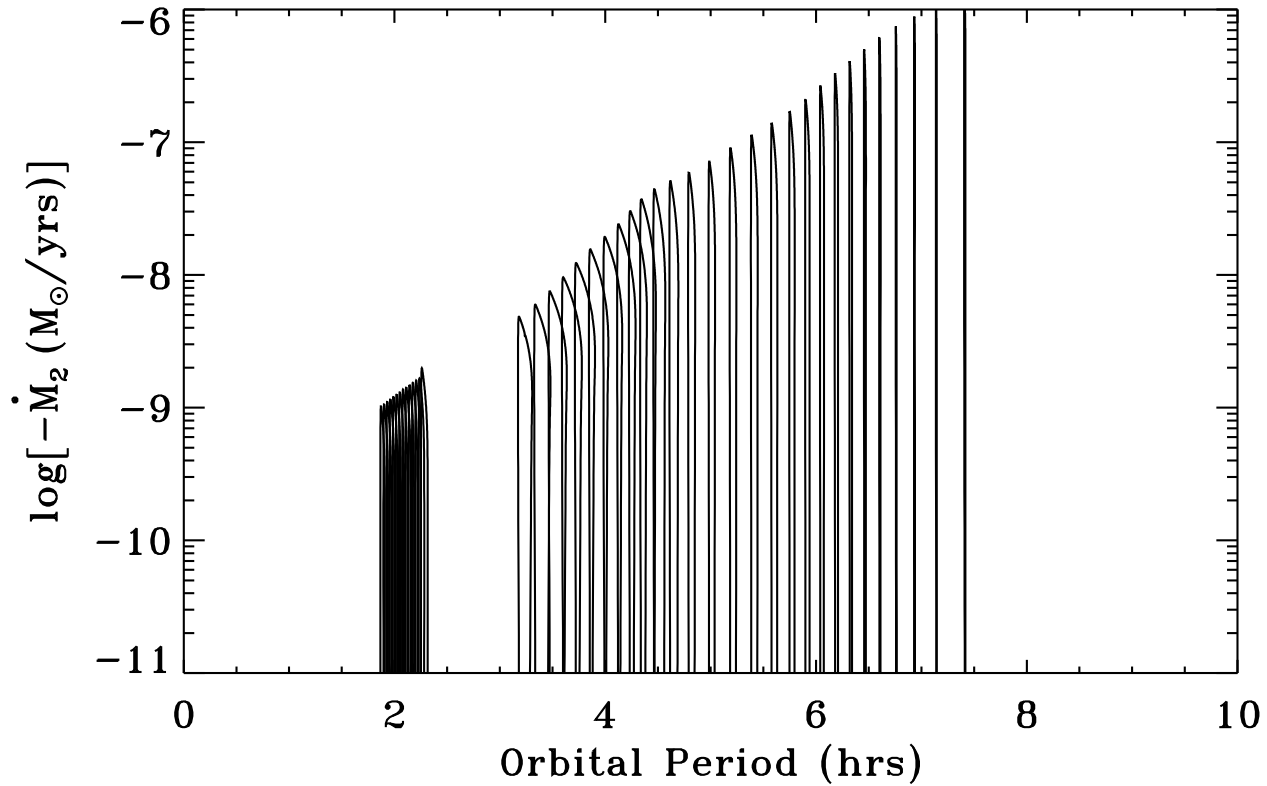}
\caption{Same evolution as Fig. 8, calculated with a  
more realistic $H_p/R_2=\epsilon= 10^{-4}$, for comparison purposes.}
\end{figure}

\begin{figure}
\plotone{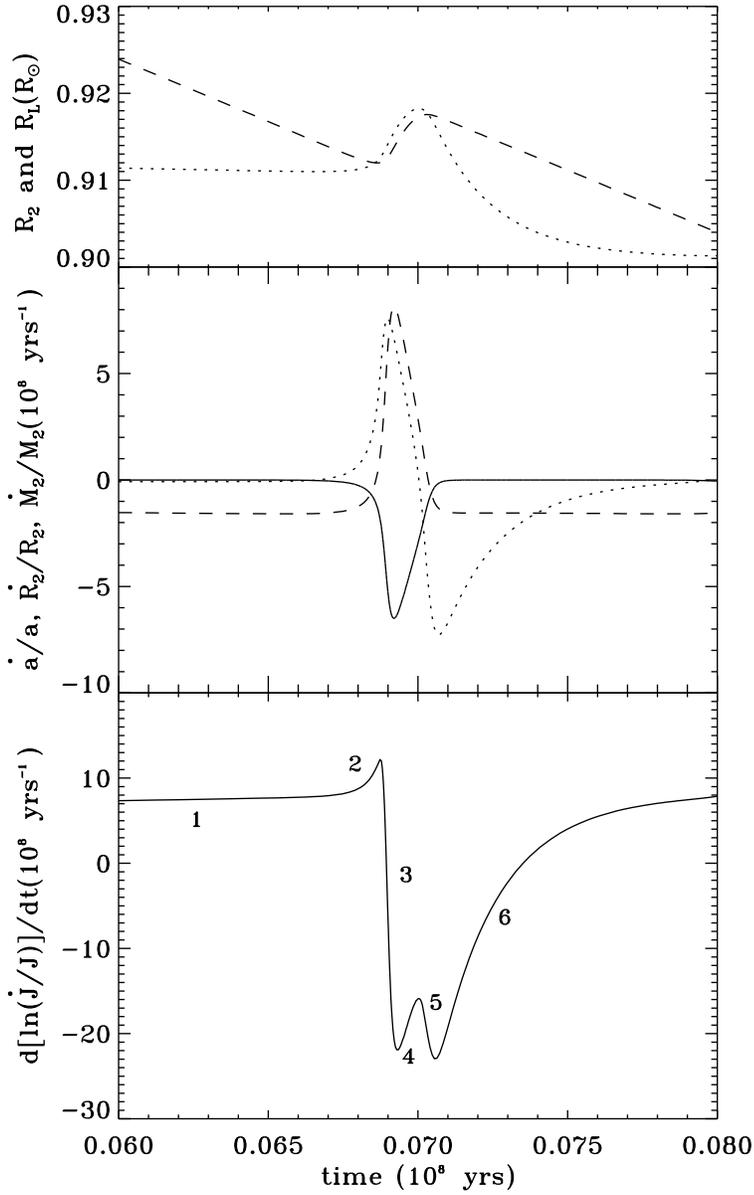}
\caption{A plot of one mass transfer cycle. In the top graph
$R_2$ (dotted) and $R_{\rm L}$ (dashed) have been plotted as functions of time. 
The middle graph shows $\dot R_2 /R_2$ (dotted), $\dot a/a$ (dashed) and 
$\dot M_2/M_2$ (solid) as functions of time. In the bottom graph  
$d[\ln{(\dot J/ J)}]/dt$ has been  plotted over the same time interval.   
This model was calculated with no CAML ($l=\beta_1=0$), with an 
irradiation efficiency of $\eta=0.1$.}
\end{figure}

\begin{figure}
\plotone{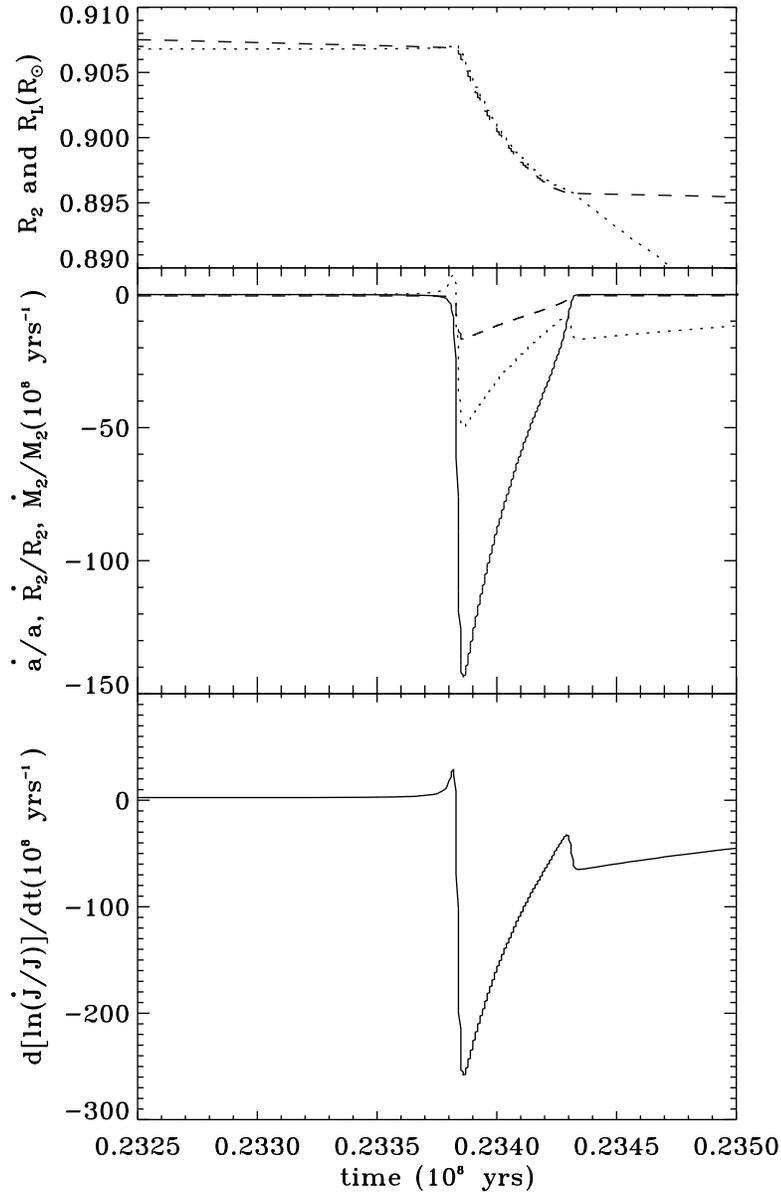}
\caption{A detailed plot of one CAML--assisted mass transfer cycle selected 
from the evolution shown on Fig. 9, at a period above 7 hours. 
In the top graph $R_2$ (dotted) and  $R_{\rm L}$ (dashed) have been plotted as 
functions of time. 
 In the middle graph $\dot R_2 /R_2$ (dotted), $\dot a/a$ (dashed) and $\dot M_2/M_2$ 
(solid) have been plotted as functions of time. In the bottom graph  
$d[\ln{(\dot J/ J)}]/dt$ has been  plotted over the same time interval.}
\end{figure}

\begin{figure}
\plotone{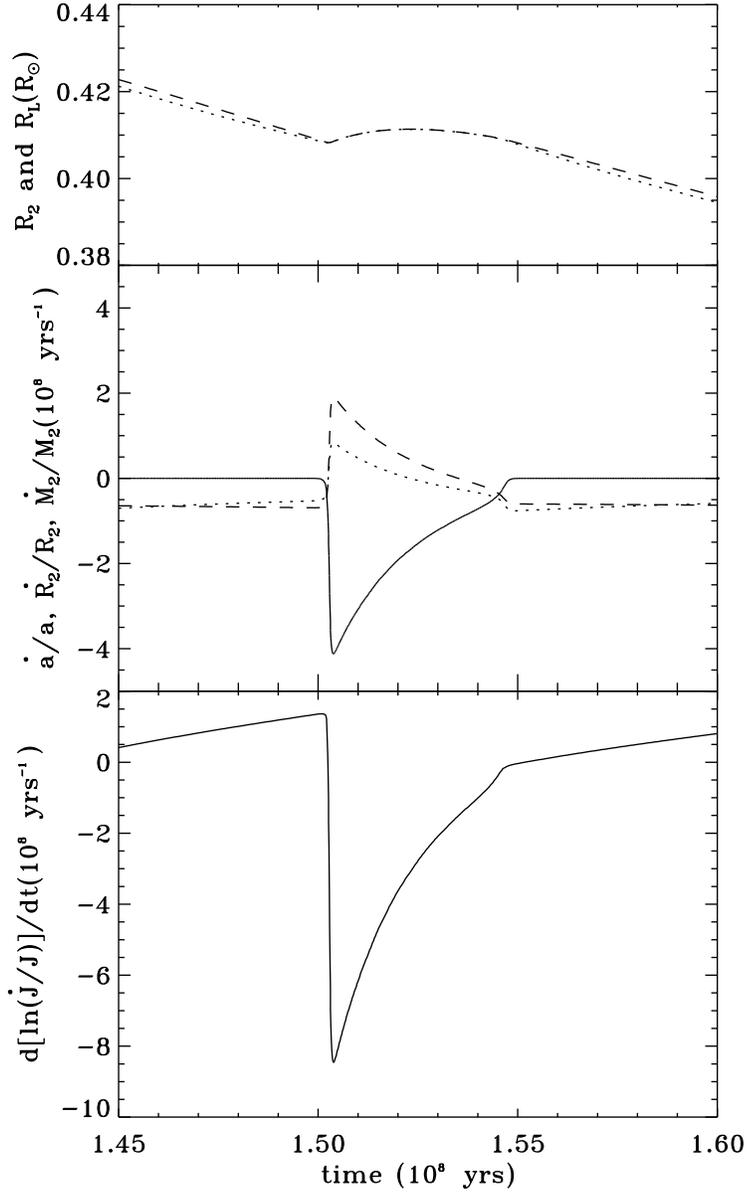}
\caption{A detailed plot of another CAML--assisted mass transfer cycle selected 
from the evolution shown on Fig. 9, at a period below 4 hours. The same conventions 
as in the preceeding two figures were used here.}
\end{figure}

\begin{figure}
 \plotone{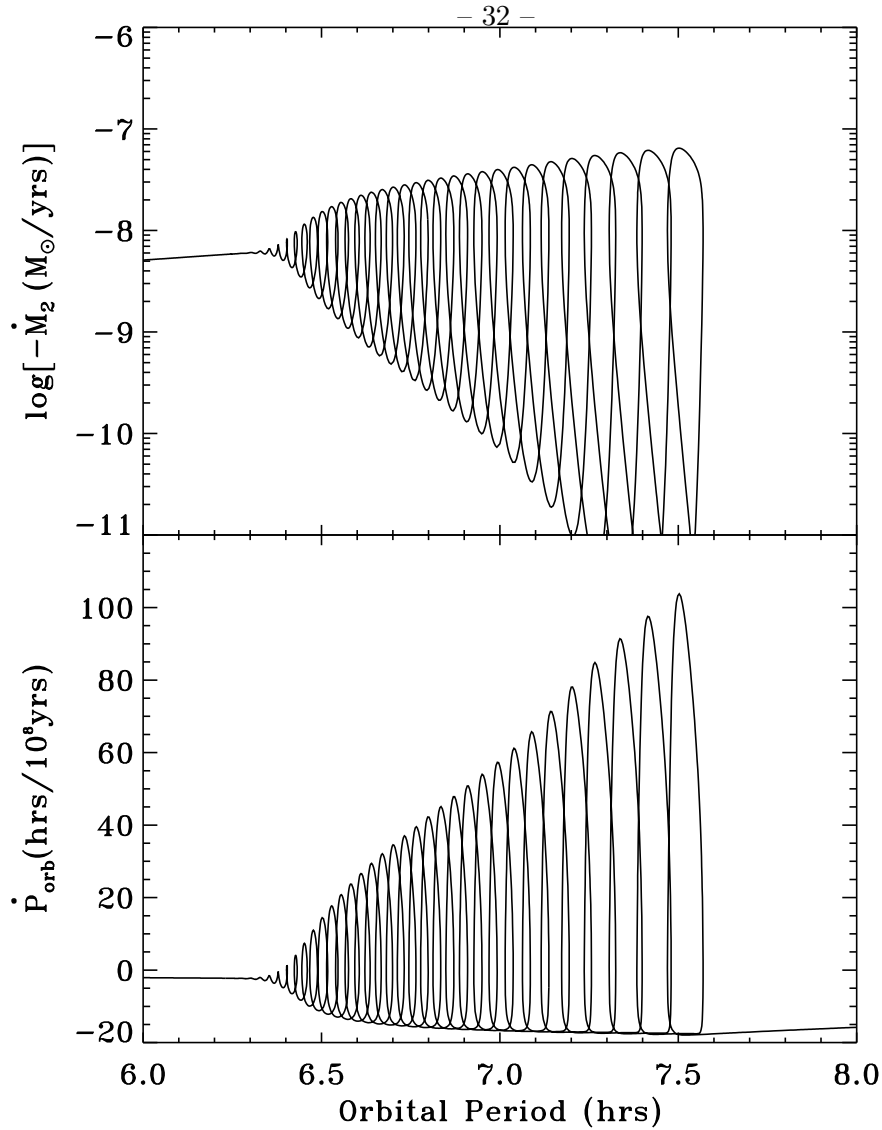}
\caption{Top: A plot of $\dot M_2$ as a function of the orbital period 
$ P_{\rm orb} $ .  Bottom: A plot
 of the orbital period derivative $\dot P_{\rm orb}$ as a 
function of orbital period.  An irradiation efficiency of $\eta=0.1$ is assumed
in this model. No  CAML is included in this simulation.
We only show a range of periods between 6-8 hrs in order to emphasize the structure 
of the oscillation. The full calculation is shown in Fig. 2 ($\eta=0.1$ panel). }
\end{figure}

\begin{figure}
\plotone{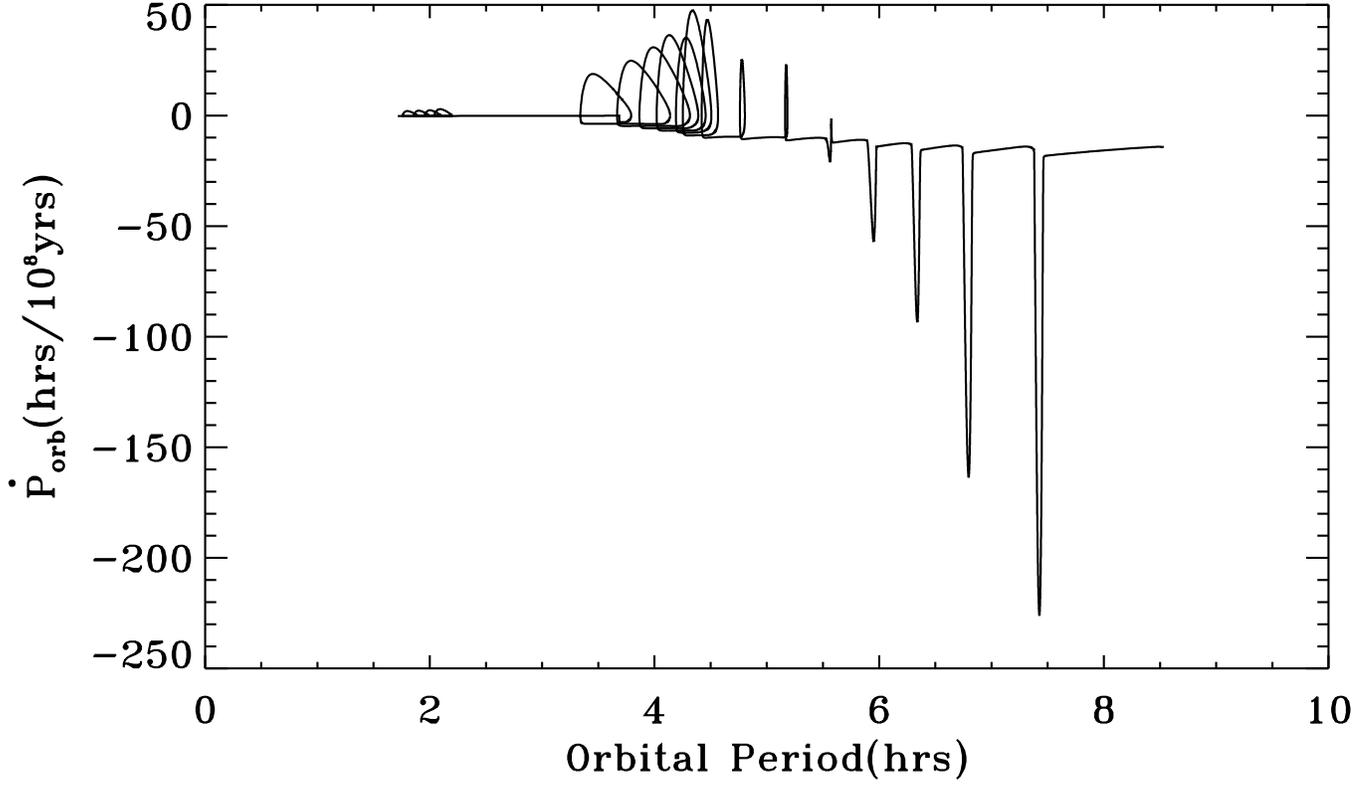}
\caption{A plot of the orbital period derivative $\dot P_{\rm orb}$ for an
evolution calculated with 
an irradiation 
efficiency $\eta=0.1$, including CAML in the form (\ref{Caml.3}) with 
$l=0.95$, and $\beta_1=0$.}
\end{figure}

\begin{figure}
\plotone{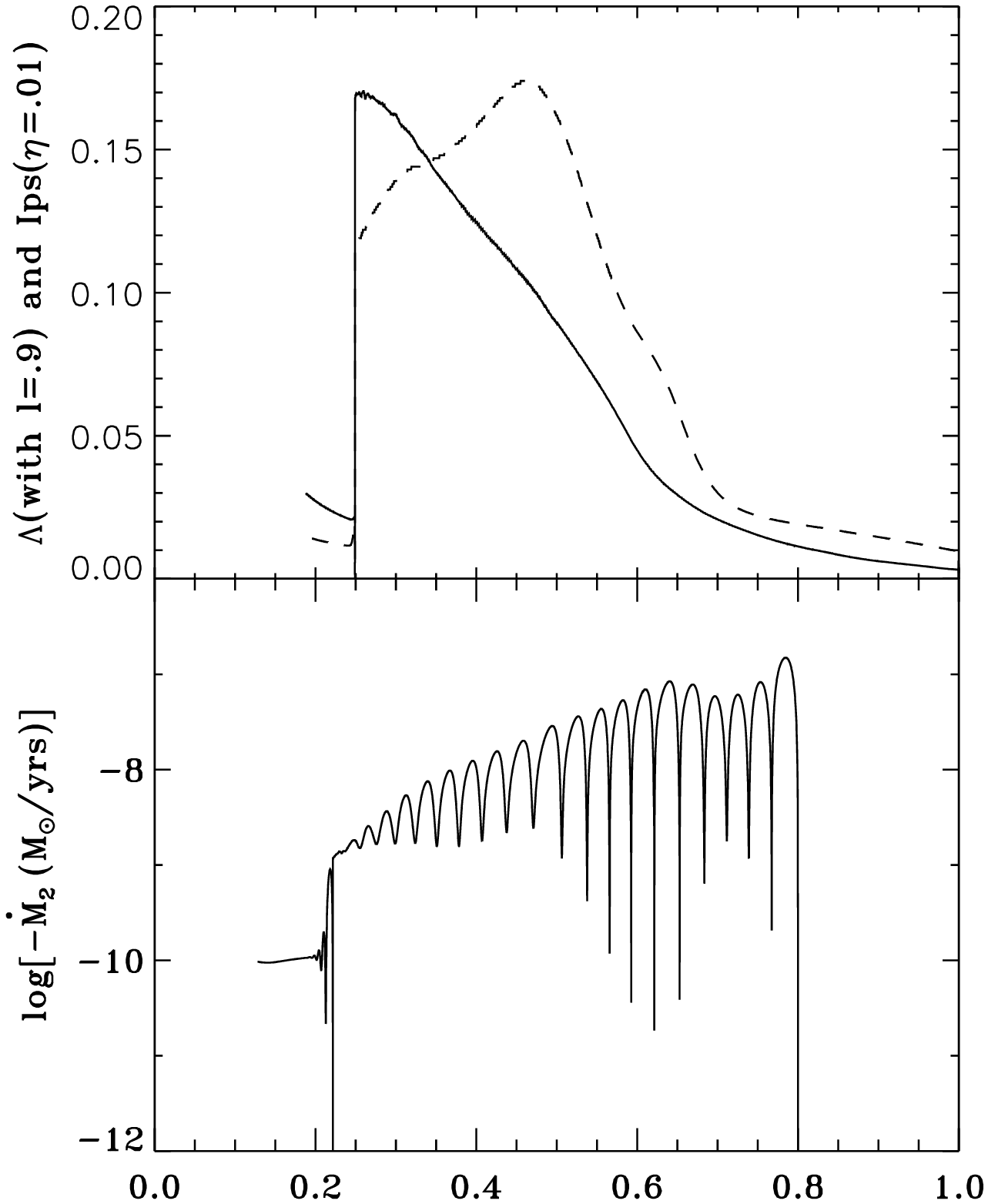}
\caption{The top panel shows the stability criterion: cycles
are expected when the dashed line is above the solid line or
$I_{\rm ps}>\Lambda$. The bottom panel shows the accretion rate as 
a function of the companion mass during an evolution with $l=0.9$. 
See text for details.}
\end{figure}

\begin{figure}
 \plotone{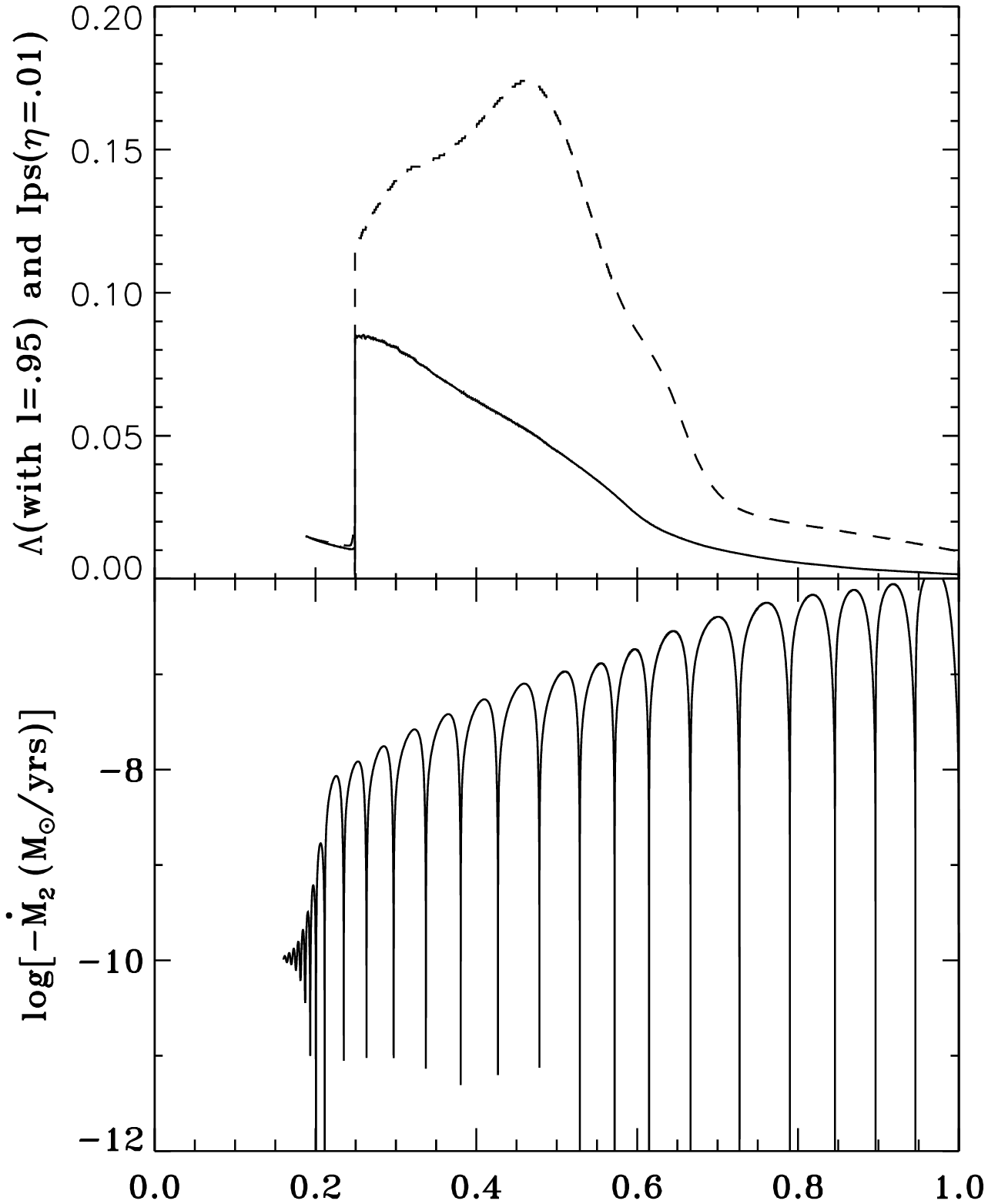}
\caption{The top panel shows the stability criterion: cycles
are expected when the dashed line is above the solid line or
$I_{\rm ps}>\Lambda$. The bottom panel shows the accretion rate as 
a function of the companion mass during an evolution with $l=0.95$. 
See text for details.}
\end{figure}

\end{document}